\documentclass[tRPC2e]{revtex4}
\usepackage{graphicx}% Include figure files
\usepackage{dcolumn}% Align table columns on decimal point
\usepackage{bm}% bold math
\usepackage{amsmath,amssymb}

\begin{document}

%\preprint{APS/123-QED}

\title{Photoionizaton of Pure and Doped Helium Nanodroplets}% Force line breaks with \\

\author{M. Mudrich}
\author{F. Stienkemeier}
%\email{Marcel.Mudrich@physik.uni-freiburg.de}
\affiliation{Physikalisches Institut, Universit\"at Freiburg, 79104 Freiburg, Germany}
\date{\today}% It is always \today, today,

\begin{abstract}
Helium nanodroplets, commonly regarded as the ``nearly ideal spectroscopic matrix'', are being actively studied for more than two decades now. While they mostly serve as cold, weakly perturbing and transparent medium for high-resolution spectroscopy of embedded molecules, their intrinsic quantum properties such as microscopic superfluidity still are subject-matter of current research. This article reviews recent work on pure and doped He nanodroplets using PI spectroscopy, an approach which has greatly advanced in the past years. While the notion of the ideal spectroscopic matrix mostly no longer holds in this context, photoionization techniques provide detailed insights into the photo-physical properties of pure and doped He nanodroplets and their relaxation dynamics following electronic excitation. Exploiting nowadays available high laser fields, even highly ionized states of matter on the nanoscale can be formed. Our particular focus lies on recent experimental progress including fs time-resolved spectroscopy, photoion and electron imaging, and novel sources of highly energetic radiation.
\end{abstract}

%Use showkeys class option if keyword
\maketitle

\section{\label{sec:Intro}Introduction}
Helium (He) nanodroplets have been a focus of research of cluster physics and physical chemistry for more than twenty years. Several review articles have been devoted to the various facets of the production, the fundamental properties, and applications of He nanodroplets~\cite{Callegari:2001,Stienkemeier:2001,Toennies:2004,Stienkemeier:2006,Choi:2006,Barranco:2006,Tiggesbaumker:2007,CallegariErnst:2011}. Today, He droplets are widely used as nanometer-sized cryo-matrices for spectroscopic studies of embedded molecules and complexes. However, their peculiar properties associated with their highly quantum nature resulting from their ultralow internal temperature of 380\,mK still pose us many riddles.

One of the useful properties of He droplets is their ability to pick-up atoms and molecules they collide with on their travel through the molecular beam apparatus. These impurities, called `dopants', are subsequently embedded into the droplet interior, or, in some cases, stick to the droplet surface. While dopant molecules are efficiently cooled internally to the droplet temperature, guest-host interactions between the molecules and the He droplet are extraordinarily weak as long as the dopants remain in their electronic ground state. This outstanding property of He droplets results from the extremely low polarizability of He as well as from the quantum liquid nature and the unique spectrum of elementary excitations of He droplets~\cite{Chin:1995,Dalfovo:1995}. Therefore, He nanodroplet isolation is a particularly well-suited technique when combined with microwave and infrared spectroscopy~\cite{Callegari:2001,Choi:2006}. Owing to the highly resolved absorption spectra of embedded molecules at very low temperature, He nanodroplets are commonly termed the ``ideal spectroscopic matrix''~\cite{Lehmann:1998,Whaley:2001}.

However, this attribute is no longer generally justified when it comes to electronic spectroscopy, where electronically excited bound states as well as the ionization continuum is probed. For larger molecules, high resolution in electronic spectra can be achieved when the overall electronic configuration does not vastly alter, and when no significant geometric changes are induced \cite{Stienkemeier:2001,Wewer:2004,Pentlehner:2010}. However, in particular for small molecules and open shell systems, upon electronic excitation or ionization, strong guest-host interactions set in which induce considerable spectroscopic line shifts and broadenings. The He droplets then turn from an inert substrate into a reactive environment which facilitates the formation of He containing neutral molecules (``exciplexes''), ionic complexes (``snowballs''), and even strongly ionized states of matter (``nanoplasmas'').

Most atoms and molecules in their electronic ground state experience a weakly attractive interaction with He. As a result, these species are located in the interior of He droplets. Electrons, however, are subject to short-range repulsion from He. As a consequence, the interaction of a dopant atom or molecule with He usually becomes repulsive upon excitation into an excited electronic orbital which is spatially more extended than the ground state. Hence, an excited dopant will be accelerated towards the surface and may eventually be expelled from the droplet. In certain cases, local dopant-He attraction and global repulsion coexist due to non-spherically symmetric dopant-He interactions. Then dopant-He$_n$, $n=1,2,\dots$ exciplexes can form as ejected free molecules. Upon ionization, strong attractive forces between the cation and He set in due to the electrostatic polarization of the He atoms surrounding the ion, which tend to drag the ion toward the droplet center to form a ion-He snowball. Once relaxed inside the He droplet, the cation again desolvates and leaves the droplet upon laser-excitation. Rydberg states of atoms and molecules, having extended electron orbits which may exceed the size of the He droplet, show the characteristics of either repulsive or attractive dopant-He interaction, depending on whether the repulsive electron-He or the attractive cation-He interaction dominates.

In these regimes of excitation or ionization of either the dopants or the He itself, a rich spectrum of new phenomena emerges which are subjects of ongoing research. New experimental approaches include ion imaging detection, time-resolved femtosecond pump-probe spectroscopy, as well as novel sources of energetic radiation using free-electron lasers and high-harmonics generation from ultrashort intense laser pulses. This overview article summarizes recent advances in the field of He nanodroplet spectroscopy involving the ionization of either the He droplet or the dopant or both by interaction with radiation. It is written from an experimentalist's perspective and, while we attempt to largely reference the relevant literature, it certainly is biased by our own research activities and personal tastes and may not cover all aspects and all the work that has been done.

After briefly reviewing the fundamentals of generating beams of He nanodroplets (Sec.~\ref{sec:Generation}), we start with a discussion of resonance-enhanced multiphoton ionization (REMPI) spectroscopy of dopants attached to He nanodroplets in Sec.~\ref{sec:Spectro}. These studies extend the work on laser-induced fluorescence spectroscopy of He droplets doped mostly with alkali metal atoms and molecules. As an additional observable, photoion mass and charge spectra (Sec.~\ref{sec:MassSpectra}) as well as photoelectron spectra (Sec.~\ref{sec:PES}) reveal direct information about the reactivity of the droplet environment following excitation or ionization of dopants. Detailed insights into the dynamics of the dopant-droplet complex initiated by laser-excitation or ionization are obtained from velocity map ion or electron imaging (Sec.~\ref{sec:IonImaging}). This technique has recently been applied to pure He droplets which are directly excited or ionized by extreme-ultraviolet (EUV) radiation. The dynamic response of excited or ionized pure or doped He droplets is directly probed by femtosecond pump-probe photoionization experiments (Sec.~\ref{sec:PP}), which are now possible even in the EUV spectral range (Sec.~\ref{sec:EUV}). In the limit of intense laser pulses, the initial ionization of dopants inside He droplets can induce ionization of the whole He droplet in an avalanche-like process thereby creating a nanoplasma, which we discuss in the last section~\ref{sec:StrongField}.

\section{\label{sec:Generation}Generation of pure and doped He droplets}
He droplets of the bosonic isotope $^4$He with sizes $\bar{N}$ ranging from a few hundred up to a few million He atoms per droplet, which are typically used for spectroscopic measurements, are generated by supersonic expansion of He gas out of a small orifice of a cryogenic nozzle into vacuum. Nowadays, both continuous and pulsed nozzles are commonly used. The process of He droplet formation is well understood and discussed in detail elsewhere~\cite{Toennies:2004,Buchenau:1990,HarmsJCP:1997}. Depending on the expansion conditions (He backing pressure $p_0$, nozzle temperature $T_0$, and nozzle diameter $d$) He droplets are generated in two different expansion regimes. In the so-called subcritical expansion (typically $T_0\gtrsim 15$~K at $p_0 = 50$~bar, $d = 5$~$\mu$m) He droplets form by condensation of the out flowing atomic He gas due to many-body collisions. During the aggregation process, the newly formed He clusters release their binding energy by evaporating He atoms. As the He density gradually drops in the course of the expansion and collisions cease, the droplet temperature levels off to $0.38$~K within about $10^{-4}$~s~\cite{Brink:1990,Hartmann:1995,Toennies:2004}. When cooling the nozzle further ($T_0\lesssim 15$~K), larger He droplets are generated by dispersing the beam of liquid He ejected out of the nozzle orifice. The distributions of He droplet sizes generated in these two regimes clearly differ from one another: subcritical expansion produces small droplets ($100\lesssim\bar{N}\lesssim 3\times 10^4$), generally referred to as ``nanodroplets'' due to their dimension in the nanometer range. Since droplet formation in this regime is a statistical process the final He droplet sizes follow a broad log-normal distribution function with a half-width comparable to the most probable size. Supercritical expansion generates droplets in the size range $3\times 10^4\lesssim\bar{N}\lesssim 10^7$~\cite{Stienkemeier:2006} having a linear-exponential size distribution \cite{Knuth:1999}. When cooling the nozzle to $T_0\lesssim 7$ K very large droplets ($10^7\lesssim\bar{N}\lesssim 10^{11}$) with velocities as low as 15 ms$^{-1}$ emerge~\cite{Grisenti:2003,Gomez:2011}. In this regime, Rayleigh instabilities break up the liquid He flow.

Absolute sizes have been measured by means of deflection of the droplets out of the beam in crossed molecular beam scattering experiments~\cite{Lewerenz:1993} or by attachment of electrons and deflection in electric fields~\cite{Knuth:1999}. Recently, it has been shown that reliable information about the mean droplet sizes can be obtained from scattering with rare gases introduced into in a scattering cell (``titration'') and by comparing characteristic mass peaks recorded using a quadrupole mass spectrometer combined with electron-impact ionization~\cite{Gomez:2011,Kornilov:2009}. In most spectroscopic experiments, where the density of isolated dopant atoms or molecules should be highest, the He droplet size is set to $3\times 10^3\lesssim\bar{N}\lesssim 3\times 10^4$ by operating at subcritical expansion conditions. Large droplets generated in the supercritical expansion regime are useful when large dopant clusters are to be aggregated inside the droplets~\cite{Tiggesbaumker:2007,Mozhayskiy:2007,Volk:2013,Denifl:2009}, or when  pure He droplets are probed directly, e.\,g. using EUV radiation~\cite{Haeften:2011,Peterka:2003,Kornilov:2011}.

Pulsed cryogenic nozzles for generating He nanodroplets were first realized by modifying conventional commercial pulsed valves~\cite{Ghazarian:2002,Slipchenko:2002,Yang:2005}. Nowadays, high-performance pulsed cryogenic nozzles are commercially available~\cite{Pentlehner:2009}. These pulsed valves are particularly advantageous for experiments using pulsed lasers operated at relatively low repetition rates ($\lesssim 1$~kHz). Compared to a continuous droplet beam, orders of magnitudes higher flux within the pulse have been reached. At the same time, the gas load is significantly reduced, which greatly relaxes the requirements for high pumping speeds at the vacuum chamber for the droplet source. Pulse lengths have been determined to fall into the range 20-–100~$\mu$s. However, when operating the nozzles at high repetition rates ($\sim$1~kHz) the increased gas load may again become a limiting factor and local heating of the nozzle due to the dissipation of electric power limits the minimal nozzle temperature $T_0$. Besides, due to velocity dispersion the temporal profile of droplet sizes within one pulse tends to be inhomogeneous~\cite{Yang:2008} which might be advantageous or disadvantageous depending on the performed experiment.

Doping of He droplets is commonly achieved by inelastic collisions inside a scattering cell, termed as the pick-up technique~\cite{Scheidemann:1990,Lewerenz:1995}. The typical length of the scattering cell is a few centimeters. At a partial pressure of molecules in the cell of about 10$^{-2}$ Pa the probability of singly doping He droplets of size $\bar{N}=5000$ is highest. Depending on the material of interest, different techniques are suitable to provide the required vapor pressure in the scattering cell. Gases or high vapor pressure liquids and solids samples are directly introduced through room temperature capillaries. Sample temperatures exceeding 1500~K have been used for evaporating metals~\cite{Reho3:2000,Ratschek:2012}. Thermal radiation from the cell or the heaters does not affect the droplet beam since the lowest dipole active transitions in He are at photon energies $\gtrsim 21$~eV. In the same way, a high temperature setup has been employed to dope with radicals by means of pyrolysis~\cite{Kuepper:2002}. In that case, an effusive continuous beam of radicals intersected the He droplet beam.

Laser evaporation has been established as an alternative way to produce doped He droplets by the authors~\cite{Claas:2003,Mudrich:2007}. The material is ablated from a rotating and translating rod by a pulsed infrared, visible or ultraviolet laser. The laser plasma is typically generated inside the He source chamber 10--20 millimetres below the droplet beam near the nozzle. Apparently the high density of He atoms accompanying the central part of the He droplet beam is needed for pre-cooling the hot atoms or molecules from the laser plasma. When using a nanosecond pulsed laser for evaporating the target material, the part of the He droplet beam with is doped arrives at the detector within a time interval of about 100~$\mu$s. In this way, doping with refractory metals as well as with fragile bio-molecules was demonstrated~\cite{Mudrich:2007}. Since a significant part of the laser-induced plasma consists of charged particles, ions are also attached to the droplets~\cite{Claas:2003}. The time-of-flight mass distributions of these ion-doped droplets revealed surprisingly large charged droplet distributions which suggest that the presence of charged particles enhances the condensation of droplets. Recently we have successfully combined a laser ablation-based doping unit with a pulsed He droplet source.

The collision energies (velocities) involved in the pick-up process are high compared to energies relevant for superfluid He (Landau critical velocity). Thus, frictionless motion is not an issue and dopants are efficiently captured by the droplets. Pick-up cross sections have been determined to be of the order of 50–-90\% of the total geometric cross section of the droplets~\cite{Lewerenz:1995}. All energy contributions released during the pick-up process (collisional, binding of dopant to droplet, internal energy of dopant, dopant-dopant binding in the case of multiple doping) are dissipated by evaporation of He atoms. As a rule of thumb, $5$~cm$^{-1}$ of energy is emitted by evaporating 1 He atom. The pick-up of dopants is a statistical process which follows the Poisson statistics in first approximation~\cite{Hartmann:1996,Toennies:2004}. More realistic models include the initial He droplet size distribution, droplet shrinking and scattering during the pick-up process, and in the case of surface-bound dopants, the desorption of dopants off the droplets by evaporation~\cite{Vongehr:2003,Vongehr:2010,Buenermann:2011}.

Most atomic and molecular species immerse into the droplet interior upon doping. Due to the generally attractive dopant-He interactions, local shell-like structures of enhanced He density around the dopants are formed. In extreme cases such as for some cations, the He density may even surpass the one of solid He, in which case the dopant-He complexes are referred to as ``snowballs''~\cite{Atkins:1959,Tiggesbaumker:2007}. The solvation energy of neutral dopants in $^4$He droplets ranges from about 50 up to about 800~K$\times k_\mathrm{B}$ for Ne and SF$_6$, respectively~\cite{Barranco:2006}, whereas that of cations reaches thousands of K$\times k_\mathrm{B}$~\cite{Buzzacchi:2001,Rossi:2004}. Here $k_\mathrm{B}$ denotes the Boltzmann constant.

Only neutral alkali and to some extent alkaline earth metals remain weakly bound to the droplet surface in dimple-like states as a result of the strong short-range repulsion of the valence electron from the surrounding He~\cite{Dalfovo:1994,Ancilotto:1995,Stienkemeier:1995,Mayol:2005}. For these species, the long-range van der Waals attraction induces shallow surface states with binding energies of about 10 K$\times k_\mathrm{B}$~\cite{Ancilotto:1995,Barranco:2006}. However, alkali clusters of sizes exceeding a critical value again submerge into the droplet interior~\cite{Lan:2011,Lan:2012,Stark:2010}. The tendency of dopants to submerge into the droplets or to stay at the surface has been rationalized in terms of balancing the energy cost of creating a void cavity to accommodate the dopant, and the gain due to the presence of He atoms in the well of the dopant-He pair potential~\cite{AncilottoJLTP:1995}.

\section{\label{sec:Spectro}REMPI spectroscopy}
The first electronic spectra of dopants attached to He droplets had been introduced using laser-induced fluorescence (LIF) detection of sodium-doped droplets \cite{Stienkemeier:1995,Stienkemeier2:1995}. LIF has proven to be the most sensitive method even for larger molecules having reasonable fluorescence quantum yields (e.g.~acenes, dyes) as compared to droplet beam depletion methods employing electron impact in combination with quadrupole mass filters or bolometers as detectors~\cite{Stienkemeier:2001}. Resonance enhanced multiphoton ionization (REMPI) as an alternative detection scheme appears appealing, since it allows to record photoionization (PI) spectra separately for every fragment ion produced in the REMPI process (dopant monomer, oligomers, dopant-He complex, mixed chemical compounds, etc.). The combination of REMPI spectroscopy with time-of-flight detection has the multiplexing advantage of measuring all fragment ion spectra at the same time. Moreover, REMPI spectroscopy does not rely on fluorescence emission by the excited dopants as LIF does. Hence, spectra and dynamics including non-radiative states can be addressed. However, ionized dopants immersed in He droplets in most cases don't end up as bare dopant ions but remain bound to the He droplets or form complexes with a small number of He atoms. In this case, although the detection of the massive ions requires particular measures, spectra on the full-sized ion-doped He nanodroplets may provide useful information (see section~\ref{sec:MassSpectra})~\cite{LoginovPRL:2011,Loginov:2012}. Alternatively, detecting photoelectron yields and spectra has proven to be a powerful technique when combined with REMPI spectroscopy (see section~\ref{sec:PES})~\cite{Loginov:2012}. Finally, the ionization step generally requires higher energy (UV) photons which makes high demands as to the required laser systems. 

A new detection scheme for He nanodroplet isolation spectroscopy using REMPI was recently demonstrated by M. Drabbels and coworkers in Lausanne~\cite{Loginov:2008}. The method relies on the complete evaporation of the droplets following excitation of dissolved molecules and the subsequent detection of the remaining unsolvated molecules by nonresonant photoionization using femtosecond laser pulses in combination with time-of-flight mass spectrometry. Thus, the method combines the advantage of background-free signal detection with that of beam depletion spectroscopy, which is a sensitive technique to detect non-fluorescing states with ultrashort lifetimes. This detection scheme has been successfully applied to recording high-resolution electronic spectra of benzene, rotaxane, and of derivatives of the nucleobase adenine~\cite{Loginov:2008,Smolarek:2009,SmolarekPCCP:2010}.

The obstacles mentioned above, which make sophisticated spectroscopies necessary, do not hold for alkali (Ak) metal dopants; that's why a large body of studies has been done for these systems. Ak metals have extremely large absorption cross sections for transitions to the lowest excited states (D$_{1,2}$-lines) which are easily accessible by near-infrared and visible lasers. In contrast to all other species which are immersed in the droplet interior upon doping, Ak dopants reside in shallow dimple states at the droplet surface. Upon electronic excitation, Ak atoms and small molecules tend to desorb off the surface of He droplets due to repulsive forces acting between the excited Ak atom or molecule and the He droplet as a whole. A simplified representation of the state-dependent interaction of Ak atoms with He nanodroplets is given by the ``pseudo-diatomic model'', in which the Ak dopant constitutes one atom and the entire He droplet the other~\cite{Stienkemeier:1996,Buenermann:2007,Lackner:2011,Loginov:2011,Callegari:2011}. This model provides a simple interpretation of the absorption spectra measured by laser-induced fluorescence emission or by REMPI and of momentum distributions of desorbed Ak atoms. The comprehensive characterization of Ak absorption spectra~\cite{Buenermann:2007} in the frame of the pseudo-diatomic model has revealed that all Ak atoms experience repulsive interaction between them and the He nanodroplets in any excited electronic state, the only exceptions being the lowest excited states of rubidium (Rb) and cesium (Cs)~\cite{Auboeck:2008,Theisen:2011}. Thus, REMPI of Ak atoms and small molecules attached to He nanodroplets generates neat atomic or molecular ions with high abundance.

\begin{figure}
\centering
\includegraphics[width=0.5\textwidth]{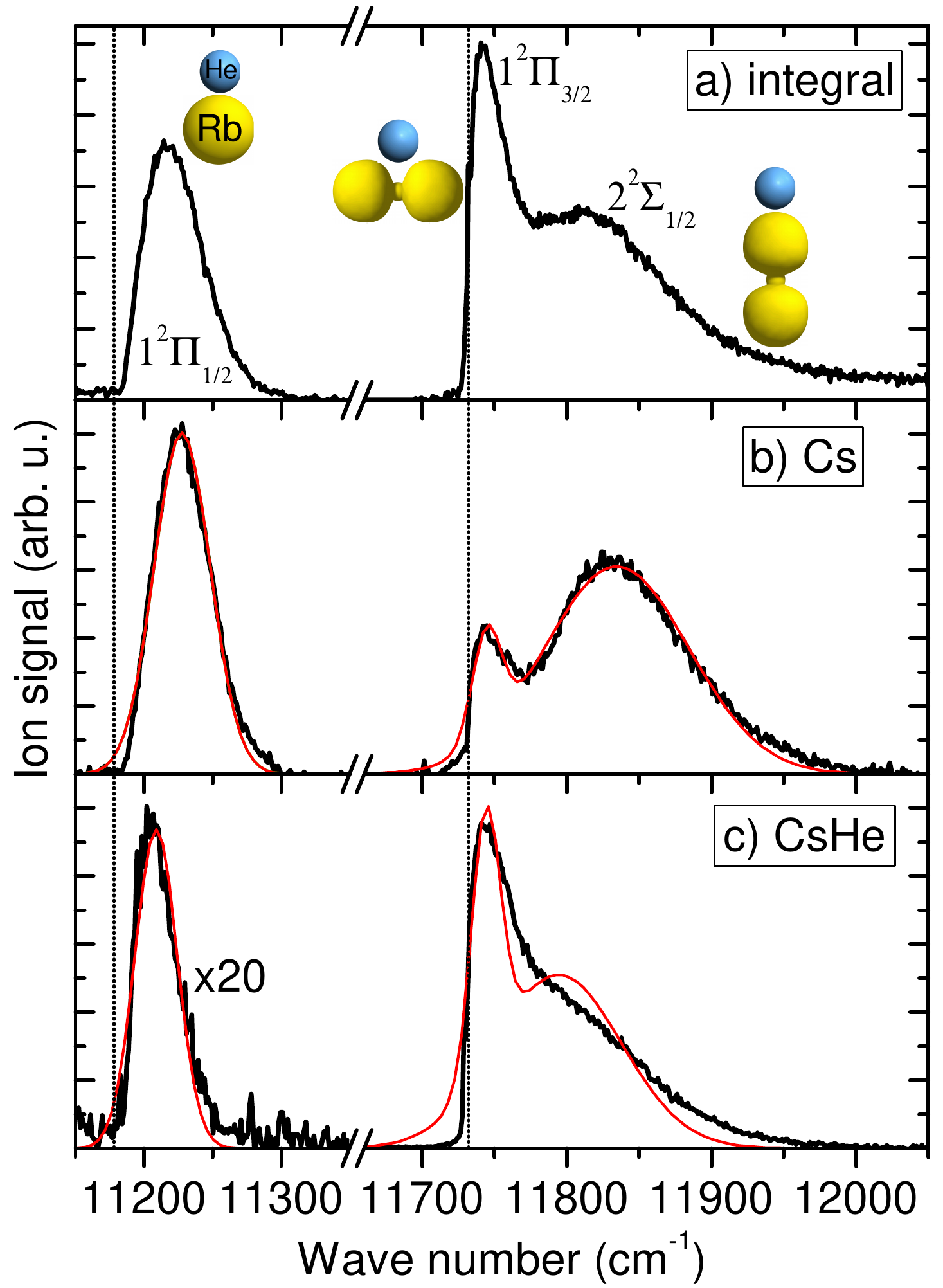}
\caption{Integral (a) and mass resolved (b, c) photoionization spectra of cesium atoms attached to He nanodroplets. The configurations of Rb and He atomic orbitals are pictorially represented in a). The thin lines in b) and c) are fits of the experimental data. Adapted from~\cite{Buenermann:2004}.}
\label{fig:CsPIspectrum}
\end{figure}
Fig.~\ref{fig:CsPIspectrum} displays the integral and product mass-resolved REMPI spectrum of Cs attached to He nanodroplets as a prototypical example for REMPI spectra of metal atoms. The vertical lines indicate the atomic D$_1$ and D$_2$-transitions. The correlating spectral features of Cs on He droplets are blue-shifted and broadened in the range $50$-$150$ cm$^{-1}$ as a result of the repulsive interaction of the excited Cs atoms with the He droplet in the Franck-Condon region. The splitting of the feature around 11800 cm$^{-1}$ into two components derives from the existence of two projections of orbital angular momentum of the atomic 6p$_{3/2}$-state with respect to the droplet surface, $1^2\Pi_{3/2}$ and $2^2\Sigma_{1/2}$. While the integral REMPI spectrum, which matches the LIF spectrum, only bears information about the absorption cross section, the spectra recorded selectively for the masses of neat Cs and of CsHe provide additional information about the response of the system to photoexcitation, the formation of CsHe complexes. These so called ``exciplexes'' are characterized by only having bound vibrational states as long as the complex is electronically excited. Upon spontaneous decay into the electronic ground state the exciplex decomposes. Accordingly, CsHe exciplexes preferentially form upon excitation of the $1^2\Pi_{3/2}$-component, which can be viewed as a Cs p-orbital lying flat on the surface such that a He atom can easily attach to the nodal plane as a result of attractive Cs-He pair-interaction.
More generally, while excited Ak atoms mostly interact repulsively with a He nanodroplet as a whole, many excited states experience pair-wise attraction between the excited Ak$^*$ atom and one individual He atom. Therefore, as the Ak$^*$ atom is repelled from the droplet surface a bound Ak$^*$He molecule or in some cases small Ak$^*$He$_n$, $n=2,3$ complexes can form simultaneously.

Alkali-He exciplexes formed on He nanodroplets have been extensively studied spectroscopically and more recently by means of time-resolved REMPI mass spectrometry (see section~\ref{sec:PP}) and ion imaging techniques (see section~\ref{sec:IonImaging})~\cite{Droppelmann:2004,Mudrich:2008,Fechner:2012,LoginovPhD:2008,Giese:2012,Loginov:2014}. AkHe exciplexes are believed to form by two different mechanisms. On the one hand, excitation of the lowest-lying p$_{1/2}$ and p$_{3/2}$-levels prepares the Ak and the nearest He atom in a state where the two atoms are initially separated by a potential energy barrier~\cite{Reho2:2000}. AkHe association can only occur by a tunneling process. Subsequent desorption of the AkHe exciplex is driven by repulsive forces acting between the exciplex and the whole droplet and possibly by vibrational relaxation of the AkHe molecule which induced its evaporation off droplet surface~\cite{Leino:2011}. On the other hand, higher-lying Ak states often feature extended attractive wells in the AkHe pair-potentials. In that case, bound vibrational states can directly be populated in a process akin to photoassociation~\cite{Fechner:2012}. These two formation schemes can be distinguished by measuring velocity distributions of the desorbing AkHe exciplex (isotropic vs.~anisotropic)~\cite{LoginovPhD:2008} or formation times (time-delayed vs.~instantaneous).

Recently, the Ak-doped He nanodroplets excited into high Rydberg states of the Ak adatom were studied by M. Drabbels and coworkers in Lausanne and by W. Ernst and coworkers in Graz. These studies were motivated by the question whether such an exotic Rydberg system consisting of a He droplet with a charged core at the center and a Rydberg electron orbiting outside the droplet can be stable. The stabilization of such a ``superatom'' could result from the $\sim 1$ eV potential barrier for electron penetration into liquid He, as theoretically predicted~\cite{Golov:1993}. However, in a more recent theoretical study, Ancilotto \textit{et al.} questioned the stability of such a system arguing that the electron orbiting outside the droplet may pull the positive ionic core close enough to the surface that fast electron-ion recombination occurs~\cite{Ancilotto:2007}.

REMPI spectra, photoelectron and ZEKE measurements revealed clearly resolved Rydberg series of broadened peaks up to principal quantum numbers $n\sim 20$ in the absorption spectra of Na, Rb and Cs attached to He droplets~\cite{LoginovPRL:2011,Lackner:2011,Lackner:2012}. While lifetimes shorter than the laser pulse ($\sim 10$ ns) were inferred for Na Rydberg states with $n\lesssim 20$, lifetimes $\tau\approx 1\:\mu$s were measured for states with $n\gtrsim 100$. Since these are still significantly shorter than the lifetimes of equivalent states in the free Na atom, these observations seem to support the suggestion that the lifetime of Rydberg states of the He droplets is governed by electron-ion recombination~\cite{Ancilotto:2007}.

All Cs Rydberg lines were found to evolve from blue-shifted for $n\lesssim 10$ to red-shifted for higher levels, indicating the transition from repulse to attractive Cs$^*$-He$_N$ interaction~\cite{Lackner:2011}. Interestingly, the transition occurs for values of $n$, where the orbital radii $R_e$ become comparable to the mean radius of the He droplets $R_{\mathrm{dr}}\approx 5$ nm. This supports the simple conception that for $R_e<R_{\mathrm{dr}}$, the Rydberg electron penetrates into the droplet and therefore the interaction is repulsive, whereas for $R_e>R_{\mathrm{dr}}$ the electron orbits the  Cs$^+$-He$_N$ cluster core and the interaction is dominated by the Cs$^+$-He$_N$ attraction.

The Rydberg line positions where found to be well reproduced by a modified Rydberg formula
\[
E_n=E_{i,\mathrm{dr}}-\frac{R_\infty}{(n-d_{\mathrm{dr}}(n))},
\]
where $E_{i,\mathrm{dr}}$ denotes the ionization energy of the CsHe$_N$ complex and $d_{\mathrm{dr}}(n)=d_{\mathrm{at}}(n)+\Delta d_{\mathrm{dr}}$ is the quantum defect consisting of the atomic term $d_{\mathrm{at}}(n)$ and an $\ell$-state dependent constant addend $\Delta d_{\mathrm{dr}}~\sim -0.1$ which accounts for the perturbation by the He droplet~\cite{LoginovPRL:2011, Lackner:2011}. $R_\infty =13.606$ eV is the Rydberg constant. The ionization energy $E_{i,\mathrm{dr}}$ was found to be lowered with respect to the gas-phase value by about 120 cm$^{-1}$ for Na. This shift matches the calculated one based on Franck-Condon factors for the ionizing transition from the neutral to the cationic NaHe$_N$ complex~\cite{LoginovPRL:2011}.

As mentioned above, only Rb and Cs excited close to their atomic D$_1$-lines remain bound to the He droplets. This has been exploited for exciting Rb and Cs into Rydberg states. Besides that, the non-desorbing lowest p$_{1/2}$-states of Rb and Cs were used as springboards for characterizing the ionization thresholds as a function of the He droplet size. Ionization energies were found to be lowered compared to the free atoms as for the Na case. The energy shift increases when going from heavy to light Ak atoms and from small to large He droplets due to the difference in polarization energies associated with the submerged Ak metal cations~\cite{TheisenJPCL:2011}. Furthermore, is was shown that large RbHe$_n$ and CsHe$_n$ snowballs are formed by resonant two-photon ionization of Rb and Cs via the lowest p$_{1/2}$-states due to the creation of Rb$^+$ and Cs$^+$ cations at the droplet surface which subsequently submerge into the droplet interior~\cite{Theisen:2010,TheisenImmersion:2011}. Recently, the alkaline-earth metal atom barium (Ba), which is thought to reside in a dimple at the droplet surface somewhat deeper than Ak atoms~\cite{Hernando:2007}, was studied using various spectroscopic techniques~\cite{Loginov:2012}. As for Ak metals, a cross-over from blue to red-shifting of the absorption lines with increasing principal quantum number was observed as well as a lowering of the ionization threshold.

Previous LIF measurements of Ak molecules formed by aggregation of atoms on the surface of He droplets have been complemented by REMPI studies of LiCs and NaCs~\cite{Mudrich:2004}, Cs$_2$~\cite{Buenermann:2004}, and Li$_2$~\cite{Lackner:2013}. Recently, absorption spectra of the mixed Ak-alkaline-earth dimer LiCa were reported~\cite{Krois:2013}. Due to the weak binding of these species to the droplets, weakly bound Ak molecules in high-spin configurations tend to be enriched~\cite{Stienkemeier:1995,Higgins:1996,Higgins:1998,Schulz:2004,Buenermann:2011}. Since the binding energy of the dopant complex is dissipated by evaporation, a high energy input tends to boil off the dopant itself thus filtering out those complexes with low binding energy.  Therefore, the mentioned studies addressed Ak dimers in triplet states. Due to their binding to the He droplets in a configuration where the molecular axis lies parallel to the surface, the coupling to the He is stronger than for singlet molecules which stand perpendicular on the droplet surface~\cite{Bovino:2009,Guillon:2011}. As a result, the vibrational lines are asymmetrically broadened towards higher energies with respect to the unperturbed transition frequencies by $\sim 100$ cm$^{-1}$. Besides, the zero-phonon line which is observed in the spectra of singlet dimers is absent in the spectra of triplet dimers~\cite{Stienkemeier:1995}. Nevertheless, the low-energy edges of the vibrational lines were found to match the expected line positions within a precision of a few cm$^{-1}$~\cite{Mudrich:2004,Lackner:2013,Krois:2013}.

The first non-alkali REMPI spectra were measured for embedded silver (Ag) clusters \cite{Federmann2:1999}. The observed spectra of Ag dimers and trimers exhibit strongly broadened and shifted transitions. In general, for larger clusters one does not expect highly resolved REMPI spectra because of the collective plasmon character of metallic clusters with corresponding lifetimes much shorter compared to the nanosecond laser pulse. Nevertheless, a narrow (56\,meV in widths) resonance of the silver octamer was observed indicating an excited state lifetime in the nanosecond range. When probing the Ag dopant monomers inside He droplets, the authors could for the first time measure highly resolved Rydberg series of Ag (principle quantum numbers $n=$20-60). Their appearance gives first evidence for the migration of excited impurities towards the surface of the He droplet within the duration of the nanosecond laser pulse~\cite{Federmann:1999}. A more detailed study of Ag atoms embedded in He nanodroplets using electron and ion imaging spectroscopy by M. Drabbels and coworkers is described in section~\ref{sec:PES}.

Recently W. Ernst and coworkers have started to investigate the formation and the deposition on surfaces of transition metal clusters aggregated inside He nanodroplets. In this context, chromium (Cr) and copper (Cu) atoms embedded in He droplets were studied by means of REMPI spectroscopy~\cite{KautschPRA:2012,Koch:2014,Lindebner:2014}. The main outcome of that work has been that absorption lines are broadened and shifted by hundreds of cm$^{-1}$ indicating that the dopants are located inside the droplets. The presence of sharp lines in the REMPI spectra showed that excited Cr and Cu atoms are ejected out of the droplets and are subsequently ionized via resonant transitions or autoionizing states within the same laser pulse. This observation is in line with previous dispersed LIF measurements revealing highly resolved emission spectra~\cite{Kautsch:2013}. As previously observed for Ak and alkaline-earth metals as well as for Ag, the He droplet environment induces fast electronic relaxation of the laser-excited Cr atoms into various low-lying levels including different spin-states. Surprisingly, the formation of the CrHe complex involving ground state He is observed, which is expected to be extremely weakly bound~\cite{Koch:2014}.

\section{\label{sec:MassSpectra}Photoion mass spectra}
A peculiar property of mass spectrometry of doped He droplet either by electron impact or by PI is the high abundance of bare dopant masses. When ionization proceeds with a large amount of excess energy the He droplets largely decompose and ions are expelled without having He attached despite of switching to the more attractive short-range ion-He interaction. Electron-impact ionization of embedded species generally proceeds as a two-step process where first a He atoms is ionized and then the charge is transfered to the dopant. Thus, a large amount of energy (3--20~eV) is deposited in the He droplet. Only when a dopant is ionized by PI close to the ionization threshold the remaining photoion tends to sink into the droplet interior. However, a conclusive understanding of the droplet response to dopant ionization is still lacking.

Fragmentation of molecular dopants upon ionization has been found not to be significantly suppressed by the He environment~\cite{Ren:2008}. However, additional doping with water molecules can efficiently buffer the fragmentation of fragile molecules. Over the last years, extensive analysis of mass spectra of doped He droplets has been carried out in the group of Paul Scheier in Innsbruck. Recent studies include larger molecules~\cite{Bartl:2013}, molecular clusters~\cite{Leidmair:2012}, mixed doping \cite{Denifl:2010} and snowball formation analysis~\cite{AnderLan:2012}.

\begin{figure}
\centering
\includegraphics[width=0.5\textwidth]{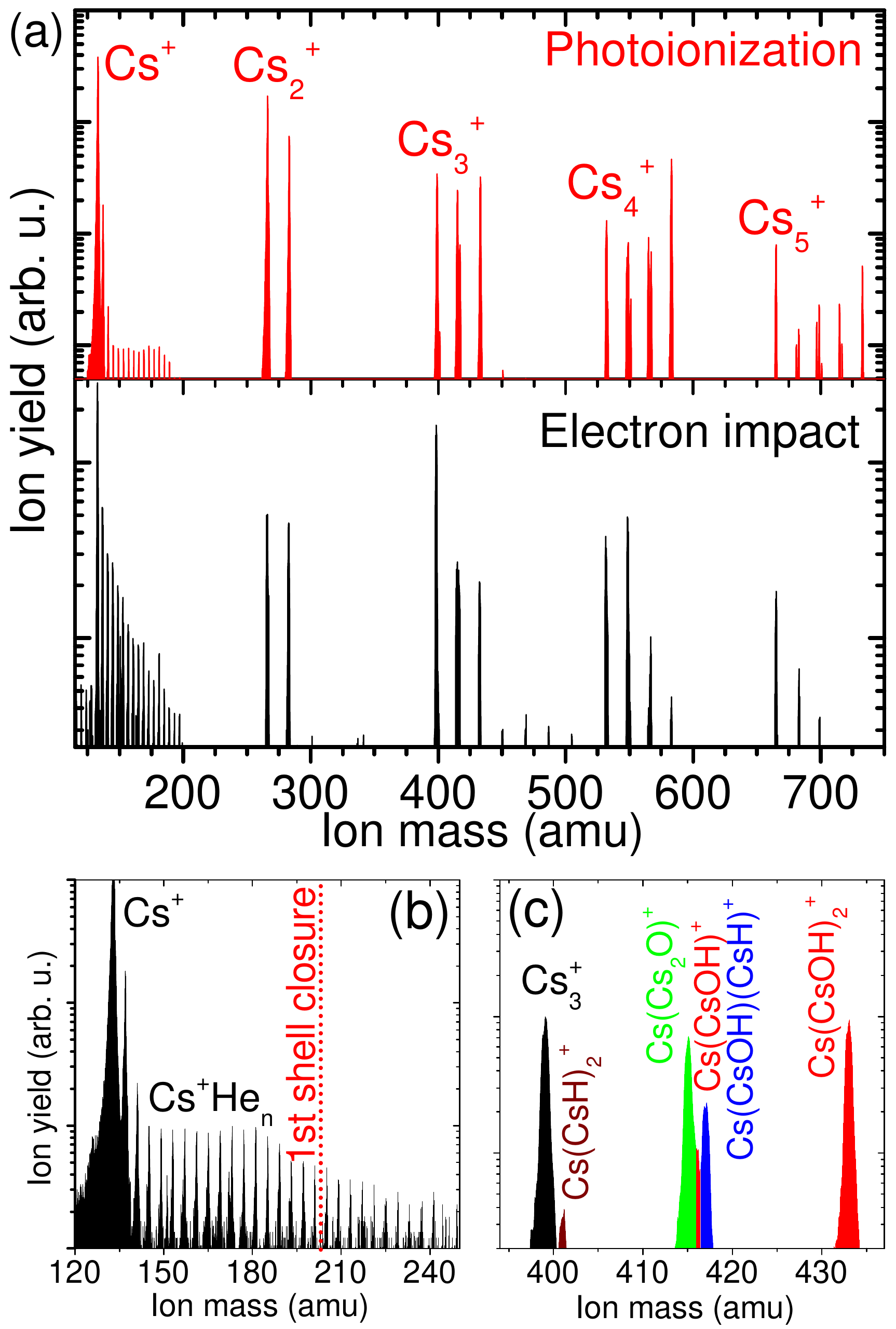}
\caption{(a) Mass spectra of mixed cesium-water clusters in He nanodroplets, recorded using femtosecond photoionization at the laser wave length $\lambda=860$\,nm (upper trace) vs. electron impact ionization (lower trace). (b) Close-up view of the mass progression of Cs$^+$He$_n$ snowball complexes. The dotted vertical line indicates the number of He atoms in the first He solvation shell around Cs$^+$ as calculated by Rossi and coworkers~\cite{Rossi:2004}. (c) Mass progression associated with compound clusters containing three Cs atoms. Adapted from~\cite{MuellerPRL:2009}.}
\label{fig:CsH2OmassSpec}
\end{figure}
Atomic ions are usually accompanied by He progressions in the mass spectra due to the formation of so-called \emph{snowballs}, shells of He atoms surrounding the ion core which have densities comparable to solid He, see Fig.~\ref{fig:CsH2OmassSpec} (a) and (b). The particular stability of the He shell at shell closures often manifests itself by characteristic steps in the mass distributions of these snowball progressions at specific magic numbers~\cite{Diederich:2005}. A detailed discussion of snowball formation and corresponding mass spectra is given in ref.~\cite{Tiggesbaumker:2007} in the context of metal clusters. The formation and stability of snowballs around Ak metal cations has been addressed theoretically~\cite{Buzzacchi:2001,Rossi:2004} and more recently experimentally~\cite{Mueller:2009,Theisen:2010,TheisenImmersion:2011}.

Furthermore, mass spectra obtained by non-resonant PI or by electron-impact ionization often contain larger dopant cluster ions as a result of the efficient aggregation of atoms and molecules into clusters when doping He droplets with more than one dopant monomer. These cluster mass progressions can extended out to masses of thousands of dopant monomers, in particular when the binding of the neutral cluster is weak thereby causing only little shrinkage of the He droplets due to cluster aggregation~\cite{Diederich:2001,Doeppner:2001,Goede:2013}. Therefore, small Ak metal clusters are presumably formed in weakly bound high-spin configurations, whereas large clusters observed in mass spectra of heavily doped large He droplets are assumed to be metallic~\cite{Schulz:2004,Droppelmann:2009,Theisen:2011}.

The appearance of neat dopant masses has even been exploited to quantify reactive processes inside He droplets by measuring the yields of product compounds in the mass spectra. For example, when co-doping Ak clusters with water, the analysis of femtosecond (fs) REMPI mass spectra revealed that Cs tends to completely chemically react with the embedded water (see Fig.~\ref{fig:CsH2OmassSpec} (a) and (c)) whereas Na preferentially forms van der Waals bound complexes under these low-temperature conditions~\cite{MuellerPRL:2009}. In addition to bare Ak clusters and to fully reacted Ak hydroxide compound clusters, various intermediate reaction products such as Ak hydrides and oxides were found in the mass spectrum (Fig.~\ref{fig:CsH2OmassSpec} (a) and (c)). These mass spectra can be qualitatively interpreted in terms of the relative abundances of Ak and water reactants in combination with the relative stability of product compounds with respect to fragmentation. Note that the mass distributions of dopant clusters and snowballs obtained by non-resonant fs PI strongly resemble those measured using electron-impact ionization (Fig.~\ref{fig:CsH2OmassSpec} (b))~\cite{MuellerPRL:2009,Tiggesbaumker:2007}. This is due to the fact that in both cases a large amount of excess energy is deposited into the droplets which induces massive fragmentation of the ions. The same holds for mass spectra obtained by direct PI of He droplets using EUV radiation, where dopant ions are formed by charge transfer ionization as in the case of electron-impact ionization~\cite{Kim:2006,Buchta:2013}.

Studies of chemical reactions inside He nanodroplets have also been carried out using electron-impact ionization as a detection method~\cite{KrasnokutskiMg:2010,KrasnokutskiSi:2010}. In terms of ion-molecule reactivity, co-doping of C$_{60}$ and water or ammonia revealed for the latter a suppression of proton transfer and the appearance of protonated ammonia ions by the C$_{60}$ aggregates~\cite{Denifl:2009,Schoebel:2011}.

Finally, in certain cases photoions submerge into the He droplet interior with high probability and ion mass distributions corresponding to nearly unfragmented ion-doped He nanodroplets can be detected. This applies to direct one-photon PI of dopant atoms~\cite{Froechtenicht:1996,LoginovPRL:2011}, to non-resonant PI using ultrashort laser pulses, to REMPI of dopant atoms, whose intermediate excited states are non-desorbing~\cite{Theisen:2010,TheisenImmersion:2011,Loginov:2007,Loginov:2012}, and to resonant REMPI of molecular dopants into the vibronic ground states of their cations~\cite{Smolarek:2010}. Note that ion-doped He droplets have also been observed when doping He droplets with preformed ions, e.\,g. using laser ablation or ion traps~\cite{Claas:2003,Mudrich:2007,Bierau:2010}. However, due to the technical difficulty in detecting these massive ions with high sensitivity, they have only rarely been used as a quantitative observable.

\section{\label{sec:PES}Photoelectron spectroscopy}
Photoelectron spectroscopy (PES), also called photoemission spectroscopy, is a well-established diagnostic tool for studying solid state samples. Since work functions in general exceed energies readily accessible with conventional lasers the most frequently used light sources are synchrotrons in the XUV and x-ray range (cf.~ultraviolet photoelectron spectroscopy -- UPS; x-ray photoelectron spectroscopy -- XPS, electron spectroscopy for chemical analysis -- ESCA). For gas-phase samples PES has proven to be a powerful spectroscopic method in particular for mass-selected ions. Apart from experiments on molecular samples, it allows to study even electronic band structures in metallic clusters~\cite{Issendorff:2005}. The applicability to He droplets depends to a large extent on the interaction of the outgoing electron with the He droplet environment which limits the achievable energy resolution.

First PES experiments of pure He droplets were carried out using synchrotron radiation by D. Neumark and coworkers in Berkeley. Various peculiarities of directly excited or ionized He droplets were identified, such as the emission of electrons having almost zero kinetic energy (ZEKEs) from pure droplets~\cite{Peterka:2003,Peterka:2007}, the indirect ionization of dopants by charge transfer or by excitation transfer out of excited and relaxed states of He, and the development of a conduction band structure in large He droplets~\cite{Wang:2008}. This work will be discussed in more detail in section~\ref{sec:EUV}.

PES of doped He droplets was first applied to Ag clusters submerged in He droplets using a magnetic bottle type spectrometer~\cite{Radcliffe:2004,Przystawik:2006,Przystawik:2007}. In such a device, all electrons from the interaction process are guided in a bottle-shaped magnetic field towards the detector. The kinetic energy of the electrons is determined by their flight times~\cite{Kruit:1983}. The appearance of a narrow feature in the PES of Ag$_8$ demonstrated that fast energy relaxation to the lower band edge of the excited $E^*$ state was present, presumably due to efficient coupling to the He droplets. However, only weak  perturbations of the PES by the interaction of the emitted photoelectrons with the He environment was observed and no evidence for a systematic shift of the ionization energy in small droplets was found. PES of Ag$_2$ again revealed fast nonradiative relaxation to be present. Both singlet and triplet states were detected.

Ag atoms embedded in He droplets were studied by M. Drabbels and coworkers using various spectroscopic techniques~\cite{Loginov:2007,Mateo:2013}. The main finding was that Ag atoms excited on the lowest electronic transitions tend to be ejected out of the droplets. Complex relaxation dynamics leads to the population of various electronic states of Ag atoms and AgHe exciplexes. The substantial deviation of the isotope ratio of the detected AgHe masses from the natural Ag abundances strongly suggests that AgHe exciplexes form by a tunneling process. PES indicated that the fraction of atoms which remain in the droplets become solvated as AgHe$_2$.

\begin{figure}
\centering
\includegraphics[width=0.7\textwidth]{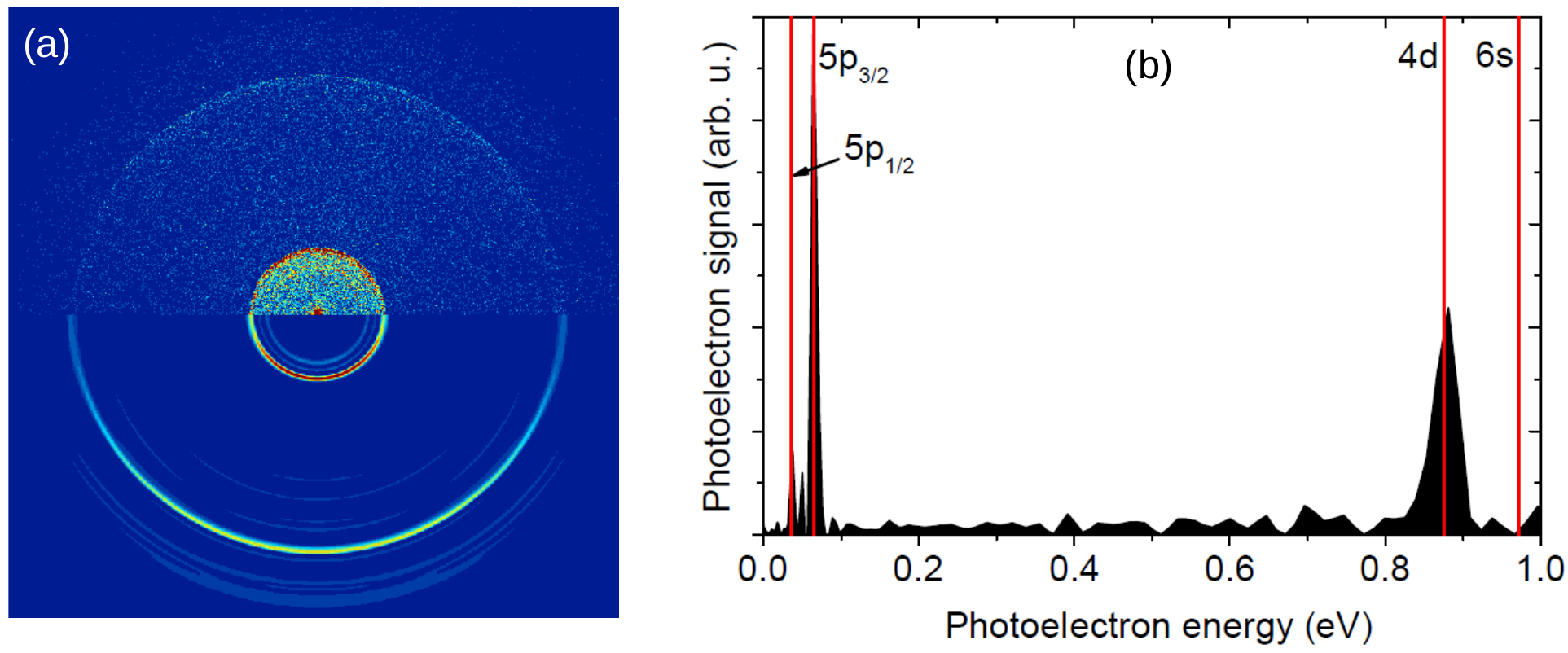}
% nur positive delays zeigen
\caption{(a) Raw (upper half) and inverse Abel transformed velocity map image of photoelectrons (lower half) recorded by exciting RbHe$_N$ complexes into the 6s$\Sigma$-state at a laser wave length $\lambda =467$ nm followed by photoionization of the desorbed Rb atom. (b) Photoelectron spectrum derived from the image. The vertical lines indicate the photoelectron energies for free Rb atoms. Adapted from~\cite{Vangerow:2014}.}
\label{fig:RbPES}
\end{figure}
Velocity map imaging (VMI) PES has recently been used to obtain complementary information about the desorption dynamics of Ak and alkaline-earth metal atoms attached to the surface of He nanodroplets~\cite{Fechner:2012,Hernando:2012,Loginov:2011,Loginov:2012}. Although the laser-induced desorption of excited Rb and Cs atoms off the He droplet surface resembles the dynamics of impulsive photodissociation of diatomic molecules (see section~\ref{sec:IonImaging}), considerable droplet-induced electronic relaxation into lower-lying atomic states is observed. As an example, Fig.~\ref{fig:RbPES} (a) depicts a raw (upper half) and inverse Abel transformed VMI of photoelectrons (lower half) recorded for Rb on He droplets excited into the 6s$\Sigma$-state of the RbHe$_N$ complex at the laser wave length $\lambda =467$ nm close to the atomic 5s$\rightarrow$6s transition~\cite{Fechner:2012}. This normally forbidden transition becomes allowed for Rb on He droplets due to the symmetry-breaking effect of the He surface. Since the atomic 6s-state is the main component of the excited 6s$\Sigma$ droplet state, one would expect to measure only the corresponding signal in the photoelectron image and spectrum. However, the image of RbHe$_N$ features three rings associated with the lower-lying 4d and 5p$_{1/2,\, 3/2}$-states. The photoelectron spectrum derived from the image is shown in Fig.~\ref{fig:RbPES} (b). Thus, despite of the surface location of Ak dopants, the coupling to the He bath in terms of electronic relaxation appears to be as strong as for other metal atoms which are embedded in the interior of the droplets~\cite{Loginov:2007,Kautsch:2013,Lindebner:2014}. In a recent study of sodium (Na) atoms excited to Rydberg states, the desorbed Na atoms were found to populate almost exclusively lower lying levels although the lowest excitation showed no droplet-induced relaxation~\cite{Loginov:2014}. The authors argued that as the principal quantum number of the excited atomic state rises, the Na-droplet interaction becomes more strongly influenced by the attraction between the He droplet and the positively charged Na core, leading to slow desorption or even solvation of the atom. As a result, transient rearrangement of the He environment during the desorption dynamics becomes important, which may induce changes in the dopant-droplet potential surfaces and the appearance of curve crossings. In addition, the energy levels are more closely spaced, which enhances the probability for curve crossing. Therefore relaxation becomes more and more likely as one approaches the ionization limit.

The influence of the He droplets on the PES was quantified by photoelectron imaging of aniline-doped He droplets by M. Drabbels and coworkers~\cite{Loginov:2005}. Compared to gas-phase spectra a blue-shift of the kinetic energy of electrons of the order of 800\,cm$^{-1}$, and tails extending 100-300\,cm$^{-1}$ to lower kinetic energies were observed. The shift was rationalized by a lowering of the ionization threshold due to polarization effects of the He environment. Within the polarizable continuum model the vertical ionization threshold in clusters, $E_{i,\mathrm{dr}}$, depends on the cluster radius $R$ as
\[
E_{i,\mathrm{dr}} (R)=E_{i,\infty} - \frac{e^2(1-\varepsilon^{-1})}{8\pi \varepsilon_0 R},
\]
where $E_{i,\infty}$ is the vertical ionization threshold in bulk He, $e$ the electron charge, $\varepsilon_0$ the permittivity of free
space, and $\varepsilon$ the dielectric constant of the cluster~\cite{Loginov:2005,Jortner:1992}. Using $\varepsilon =1.055$ for bulk liquid He, one obtains a typical He droplet size-dependent shift $E_{i,\mathrm{dr}} - E_{i,\infty} \approx -100$ cm$^{-1}$ for droplets containing $3000$ He atoms which have a radius $R_{\mathrm{dr}}\approx 3.2$\,nm. The tails of the peak in the photoelectron spectra are indicative of electronic relaxation. Furthermore, the spectra revealed the depletion of low kinetic energy electrons at larger droplet sizes, possibly due to the localization of the slow electrons in the larger droplets followed by recombination with the aniline ion. VMI-PES studies of pure and doped He droplets using direct PI by EUV radiation were performed by D. Neumark, O. Gessner and coworkers in Berkeley and more recently by our group. These will be reviewed in section~\ref{sec:EUV}.

%The polarization model does not hold for faster electrons probed in synchrotron experiments at pure He droplets \cite{Peterka:2007}. Here, shifts up to 0.5~eV could be rationalized including covalent attraction of He neighbors in a Franck-Condon simulation. These experiments were later extended to doped (Kr, Xe, SF$_6$) droplets further discussing PES and energy loss mechanisms  \cite{Peterka:2006,Wang:2008}.

\section{\label{sec:IonImaging}Ion imaging}
Imaging the velocity distribution of ions created by PI has been established as one of the key detection techniques in the field of molecular reaction dynamics~\cite{Whitaker}. In the simplest realization, photoions or electrons are imaged onto a position sensitive detector (micro channel plate in combination with a phosphor screen and a camera or a delay line detector) according to their transverse velocity with respect to the spectrometer axis. This is achieved by using a configuration of three electrodes -- one repeller and two bored extractor plates which act as an electrostatic lens~\cite{Eppink:1997}. The full tree-dimensional angular velocity distribution of the emitted ions or electrons can be recovered if cylindrical symmetry with the axis pointing perpendicular to the spectrometer axis is provided. In PI experiments where the order of the relevant excitation or ionization process is well-defined (one or two-photon transition), this is realized by aligning the laser polarization perpendicular to the spectrometer axis. In this case, the measured two-dimensional projection can be inverse-Abel transformed using various algorithms~\cite{Vrakking:2001,Garcia:2004} and both velocity (kinetic energy) spectra and as well as angular distributions are obtained. In the case of electron detection, these distributions give detailed insight into the electronic state which is photoionized. VMI detection of photoions is particularly useful for experiments where laser excitation induces photodissociation and the photofragments acquire substantial kinetic energy. Ion VMI is usually combined with mass gating by switching on the detector within a short time interval which corresponds to the arrival time of a specific ion mass due to time-of-flight mass dispersion. Extensions of the VMI technique are three-dimensional ion imaging by simultaneously measuring ion positions and flight times~\cite{Chichinin:2002}, as well as slice imaging by selecting parts of the ion cloud through fast gating of the detector~\cite{Townsend:2004}. An even higher level of detail is reached  when photoions and photoelectrons are detected in coincidence (PEPICO~\cite{Baer:1991,Buchta:2013,BuchtaJCP:2013}).

VMI of photoions was first applied to doped He nanodroplets by M. Drabbels and coworkers for studying the translational motion of neutral molecular dissociation fragments through the He droplet~\cite{Braun:2004}. Due to the high speeds of the photofragments, the interaction with the He was determined by binary collisions instead of revealing the quantum properties of the superfluid He droplets.

Upon electronic excitation, open-shell atomic and molecular dopants tend to be ejected out of the He droplets due to the repulsive interaction of the excited valence electron with the surrounding He. Alkali metal atoms, which are loosely bound at the droplet surface due to their extended valence electron shell already in the ground state, are prototypical examples of such species that experience strong repulsion upon electronic excitation. Thus, the excited Ak atoms desorb off the He droplet surface akin to the photodissociation of a diatomic molecule, where the whole He droplet acts as one single atom (pseudo-diatomic model).

\begin{figure}
\centering
\includegraphics[width=0.7\textwidth]{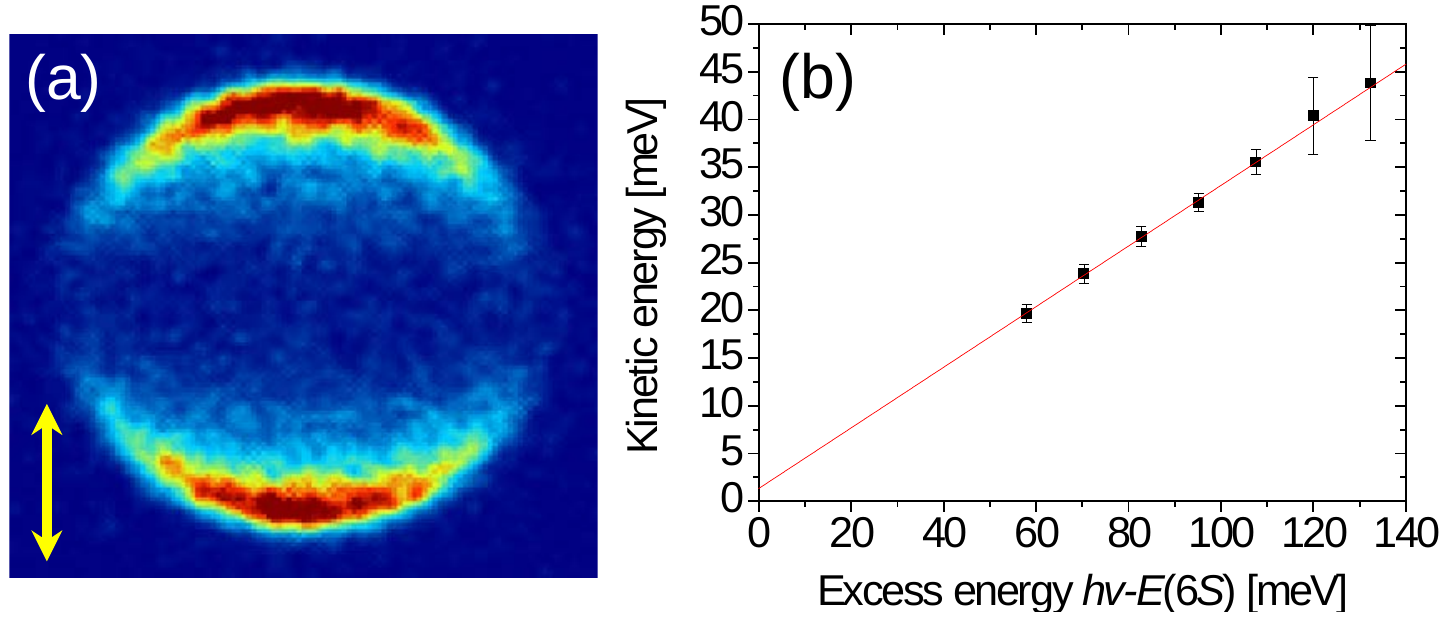}
\caption{(a) Velocity map ion image of Rb$^+$ recorded by photoexcitation to the 6s$\Sigma$-state and subsequent photoionization at $\lambda=480$ nm. (b) Rb$^+$ mean ion kinetic energies inferred from ion images measured at various values of the laser wave number within the 6s$\Sigma$ absorption band of the RbHe$_N$ complex. Adapted from~\cite{Vangerow:2014}.}
\label{fig:IonImageRb6s}
\end{figure}
Fig.~\ref{fig:IonImageRb6s} (a) displays a typical raw VMI image recorded upon excitation of the $^2\Sigma_{1/2}$-state of the RbHe$_N$ complex which correlates to the 6s atomic state of Rb. The pronounced anisotropy of the ion velocity distribution with respect to the laser polarization (double-sided arrow) and the sharp ring structure confirm the picture of pseudo-diatomic dissociation of the RbHe$_N$ complex initiated by the excitation of a parallel $\Sigma$-$\Sigma$-transition in this case. In general, the analysis of the angular anisotropy of photoion distributions provides valuable information about the symmetries of the involved excited states.

Surprisingly, the mean kinetic energy release $\langle E_{\mathrm{kin}}\rangle$ of the Rb fragment very closely follows a proportionality with respect to the excess energy given by the difference between photon energy and the atomic level energy, $E_{\mathrm{ex}}$, see Fig.~\ref{fig:IonImageRb6s} (b), although a rich spectrum of internal modes is excited in the He droplets according to time-dependent density functional calculations~\cite{Hernando:2012,Fechner:2012,Vangerow:2014}. From the slope $\eta=\Delta \langle E_{\mathrm{kin}}\rangle / \Delta E_{\mathrm{ex}}$ of the linear fit in Fig.~\ref{fig:IonImageRb6s} (b) we can infer the mass of that part of the He droplet which effectively interacts with the excited Ak atom in the desorption reaction,
\[
m^{\mathrm{Ak}}_{\mathrm{dr,eff}}=m_{\mathrm{Ak}}\frac{\eta}{1-\eta},
\]
where $m_{\mathrm{Ak}}$ denotes the mass of the Ak atom. For the Ak species Li, Na, Rb and Cs excited to the lowest excited s$\Sigma$-states the following values of $m_{\mathrm{dr,eff}}$ in units of He atomic masses have been found: $m^{\mathrm{Li}}_{\mathrm{dr,eff}}\approx 4$, $m^{\mathrm{Na}}_{\mathrm{dr,eff}}\approx 6$, $m^{\mathrm{Rb}}_{\mathrm{dr,eff}}\approx 10$, and $m^{\mathrm{Cs}}_{\mathrm{dr,eff}}\approx 13$~\cite{Hernando:2012,Vangerow:2014}. Since the linear fit function extrapolates toward the energy of the free atomic level at $E_{\mathrm{ex}}=0$, this kind of measurement can be used for identifying the atomic state from which the studied droplet-state derives~\cite{Loginov:2012,Loginov:2014}.

%The desorption dynamics of NaHe$_n$ exciplexes is found to evolve from direct ejection due to NaHe$_n$-droplet repulsion to thermal evaporation in the case of excitation into high Rydberg states  $n\gteq 10$.

A similar behavior in terms of the ejection out of the He droplets upon electronic excitation was found for the surface-bound alkaline-earth metal atom Ba~\cite{Loginov:2012}. Even the transition metal atoms Ag~\cite{Loginov:2007,Mateo:2013}, Cr and Cu~\cite{KautschPRA:2012,Koch:2014,Lindebner:2014}, which are initially submerged in the He droplet interior, are found to partly desolvate upon excitation and to leave the droplets as free atoms (see section~\ref{sec:Spectro}).

The tendency of dopants to be expelled out of the droplets upon optical excitation even if initially submerged in the droplet interior was exploited in recent experiments for determining the final speed distributions of the ejected particles for a variety of atomic and molecular species. These measurements in combination with time-dependent density functional calculations clearly revealed the existence of a critical Landau velocity for the undamped motion in superfluid He even for nanodroplets consisting of only a thousand He atoms~\cite{Brauer:2013,Mateo:2013}. This adds one more fine demonstration of the superfluid nature of these confined systems.

These studies have been further extended to atomic and molecular ions, which also tend to be ejected out of the He droplets upon electronic or vibrational excitation, respectively~\cite{Zhang:2012,Smolarek:2010,Brauer:2011}. Surprisingly, the dynamics following the
infrared excitation of a molecular ion is governed by a non-thermal process in which the ion is ejected from the He droplet instead of being cooled by fast energy transfer to the droplets and evaporation of He atoms. This peculiar behavior of He droplets was discovered by M. Drabbels and coworkers and is currently being further investigated. Moreover, it was found that the effect of the He environment on the absorption spectra, in terms of matrix shift and line broadening, is quite similar for ions and neutrals despite the fact that ions interact much more strongly with He than neutrals. Thus, infrared and also optical excitation of molecular ions detected by the ejection of the ions from the He droplets provides a novel, highly sensitive spectroscopic method for cold molecular ions~\cite{ZhangJCP:2012}.

In recent experiments carried out in the group of H. Stapelfeldt in Aarhus, VMI was combined with a fs pump and picosecond (ps) probe PI scheme to probe the rotational dynamics of molecules embedded in He nanodroplets. These will be discussed in section~\ref{sec:PP}. Ion imaging has also been applied to pure and doped He nanodroplets ionized by EUV radiation from a synchrotron and from high-harmonics generation from intense fs laser pulses by D. Neumark, O. Gessner and coworkers in Berkeley, see section~\ref{sec:EUV}.

\begin{figure}
\centering
\includegraphics[width=0.5\textwidth]{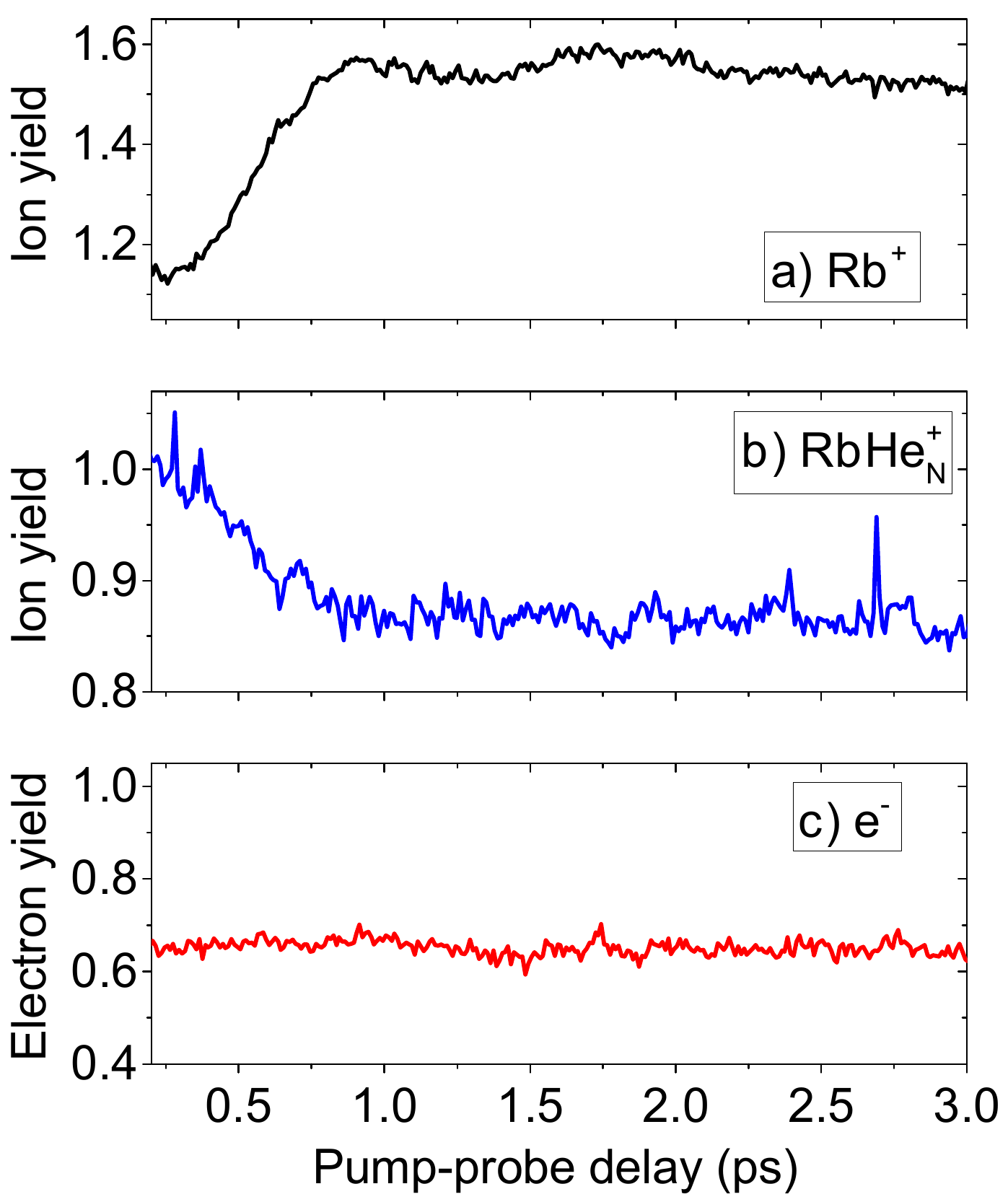}
\caption{Transient yields of Rb$^+$, Rb$_n^+$ ($n>500$), and electrons recorded upon one-color pump-probe resonant ionization of Rb-doped He nanodroplets via the 6p$\Sigma$-state at $\lambda = 400$ nm.}
\label{fig:RbDesorption}
\end{figure}
\section{\label{sec:PP}Time-resolved photoionization}
Alkali metal atoms attached to the surface of He droplets have naturally been the first systems to be studied by fs spectroscopy. Owing to the large absorption cross sections for the lowest transitions, which fall into the tuning range of the Ti:Sapphire laser in the cases of K, Rb and Cs, REMPI studies are possible even using the direct output of a standard Ti:Sapphire oscillator without pulse amplification. Furthermore, the tendency of Ak atoms to desorb off the surface of He droplets upon resonant excitation facilitates the sensitive detection of free ions generated by REMPI. Thus, by varying the delay time between a first ``pump'' laser pulse which resonantly excites a droplet-bound Ak atom and a second ``probe'' pulse which ionizes the atom, the desorption process can be followed in real time. Fig.~\ref{fig:RbDesorption} shows a typical pump-probe measurement of the transient yield of Rb$^+$ ions generated by REMPI via the perturbed 6p$\Sigma$-state of the RbHe$_N$ complex at $\lambda = 400$ nm. The increase of the Rb$^+$ yield at delay times $300<\tau<900$~fs is interpreted as the manifestation of the desorption process. At short delay $\tau\lesssim 500$~fs, ionization of Rb at short distance from the He surface induces the sinking of the Rb$^+$ ion into the droplet under the influence of strong polarization forces. Therefore the abundance of the detected neat Rb$^+$ ions is reduced and the yield of large RbHe$^+_N$ ($N>500$) complexes is enhanced at the same time. However, the yield of photoelectrons, which is indicative for the total rate of PI events, remains constant at delay times exceeding the temporal overlap of pump and probe laser pulses at $\tau>150$~fs. Thus, PI of Rb in the proximity of the He droplet surface appears to be as efficient as for the free atom after its desorption.

In spite of the repulsion of excited Ak atoms from He droplets for most of the excited states, pair-wise attraction between Ak$^*$ and individual He atoms can induce the formation of Ak$^*$He$_n$, $n=1,2,3$ exciplexes as a competing process to desorption of Ak$^*$ off the droplet surface. While the equilibrium properties of Ak$^*$He exciplexes are now well characterized even including an environment given by a He cluster or film~\cite{Leino:2011}, the dynamics of the formation process of Ak$^*$He exciplexes still eludes from an accurate description~\cite{Droppelmann:2004,Mudrich:2008,Giese:2012}.

\begin{figure}
\centering
\includegraphics[width=0.5\textwidth]{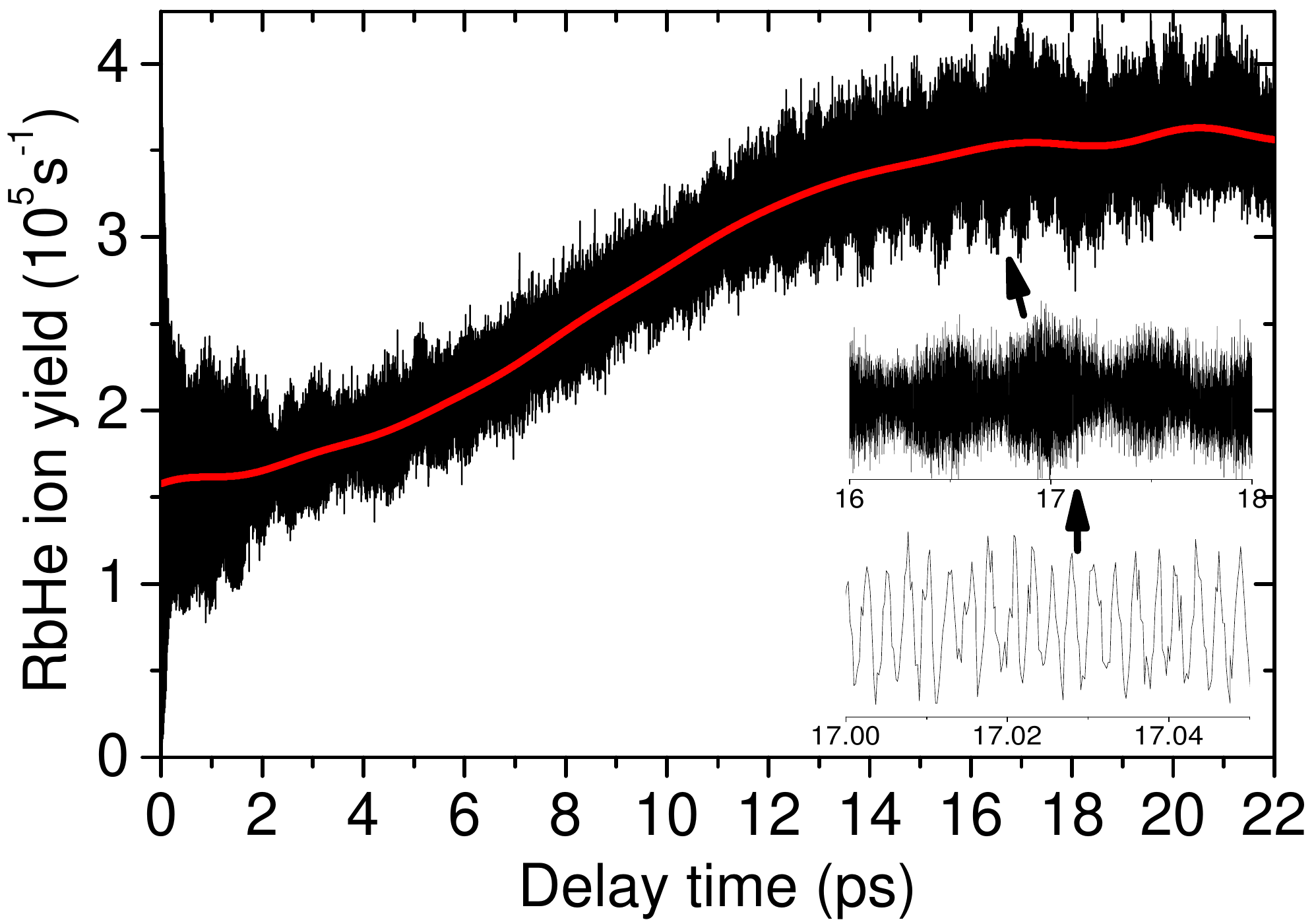}
\caption{Yield of ionized RbHe exciplexes created by photoexcitation of Rb atoms attached to the surface of He nanodroplets at $\lambda=780$ nm as a function of the delay between femtosecond pump and probe pulses. Adapted from~\cite{Mudrich:2008}.}
\label{fig:RbExciplex}
\end{figure}
The dynamics of the formation of sodium (Na) and potassium (K) exciplexes (Na$^*$He, K$^*$He) has first been studied by time-resolved fluorescence emission spectroscopy, yielding formation times in the range of tens of ps~\cite{Reho2:2000}. More recently, the K$^*$He and Rb$^*$He exciplex formation dynamics was probed using fs pump-probe spectroscopy which revealed a Rb$^*$He signal rise time of 8.5~ps, see Fig.~\ref{fig:RbExciplex}~\cite{Schulz:2001,Droppelmann:2004,Mudrich:2008}. The fast modulation of the data results from quantum interference and contains information about the REMPI spectrum in the spectral range which falls into of bandwidth of the fs laser~\cite{Mudrich:2008}. Theoretical models including one-dimensional semiclassical tunneling~\cite{Reho2:2000}, quantum-classical modeling~\cite{Pacheco:2007}, semiclassical path integral molecular dynamics~\cite{Takayanagi:2004}, and quantum Monte Carlo approaches~\cite{Leino:2011} have predicted values for the exciplex formation times ranging from 1.7 ps for the lithium-He exciplex Li*He, to 31 ps for Rb*He in the $1^2\Pi_{3/2}$-state. The role of the superfluid properties of He on the exciplex formation times has been directly probed in experiments with $^3$He droplets \cite{Droppelmann:2004}. Surprisingly, the formation of Rb$^3$He takes longer compared to the heavier isotope Rb$^4$He. This rules out the tunneling process from being the bottleneck of RbHe exciplex formation.

The dynamics of exciplex formation on He nanodroplets can be viewed as a two-step process. It is initiated by the electronic excitation of the Ak adatom which induces the attachment of a He atom directly by populating a bound molecular state or by tunneling. The excitation represents a sudden perturbation of the relaxed AkHe$_N$ doped droplet complex. In response, the local He environment of the excited Ak$^*$ atom rearranges within about 1~ps owing to the competing repulsion of the excited Ak$^*$ with respect to the whole droplet and the attractive Ak$^*$-He pair-interaction. This fast dynamics leads to a non-thermal distribution of populations of bound vibrational states of the Ak$^*$He molecule peaked at intermediate levels through dissipative coupling to the He droplet, as inferred from the quantum interference measurements of Rb$^*$He (Fig.~\ref{fig:RbExciplex})~\cite{Mudrich:2008}. A recent picosecond spectroscopic study has evidenced a second phase of vibrational relaxation proceeding on a much longer time scale~\cite{Giese:2012}. For Rb$^*$He in the $1^2\Pi_{3/2}$-state the vibrational population continued to relax towards the ground state even after delay times as long as 1.7~ns, which implies that at least part of the exciplexes remains bound to the droplets, as suggested by recent quantum Monte Carlo simulations~\cite{Leino:2011}.

Owing to the high mobility of the dopants inside as well as on the surface of He droplets, Ak dimers and trimers readily form when doping the droplets with on average two or more Ak atoms. The weak binding of Ak species to the droplets leads to an enrichment of weakly bound high-spin configurations, that is the lowest triplet $1^3\Sigma_u^+$-state of homonuclear dimers and the lowest quartet $1^4A_2'$-state of homonuclear trimers~\cite{Higgins:1996,Higgins:1998}. For these species, we have observed surprisingly long-lived coherent vibrational wave-packet dynamics by means of fs pump-probe REMPI spectroscopy~\cite{Claas:2007,Mudrich:2009,Giese:2011}. In the case of K$_2$, vibrational wave packet motion was also measured in the singlet manifold~\cite{Claas:2006}.
The vibrational transients measured with K$_2$ showed a clear signature of couplings of the K$_2$ vibronic motion to the He environment, such as damping of the oscillation amplitude, transient shifting of the oscillation frequency, transient appearance and disappearance of vibrations in different electronic states~\cite{Claas:2006}. The characteristic time constants for these effects fall into the range 3-8~ps. Detailed quantum dynamics simulations showed remarkable agreement with the experimental findings provided that dissipative coupling of K$_2$ to the He bath as well as desorption off the droplets were taken into account. These model calculations even allowed us to speculate on the influence of superfluidity on the microscopic vibrational dynamics of a single molecule, for which an accurate theoretical description is still missing~\cite{Schlesinger:2010}.

\begin{figure}
\centering
\includegraphics[width=0.5\textwidth]{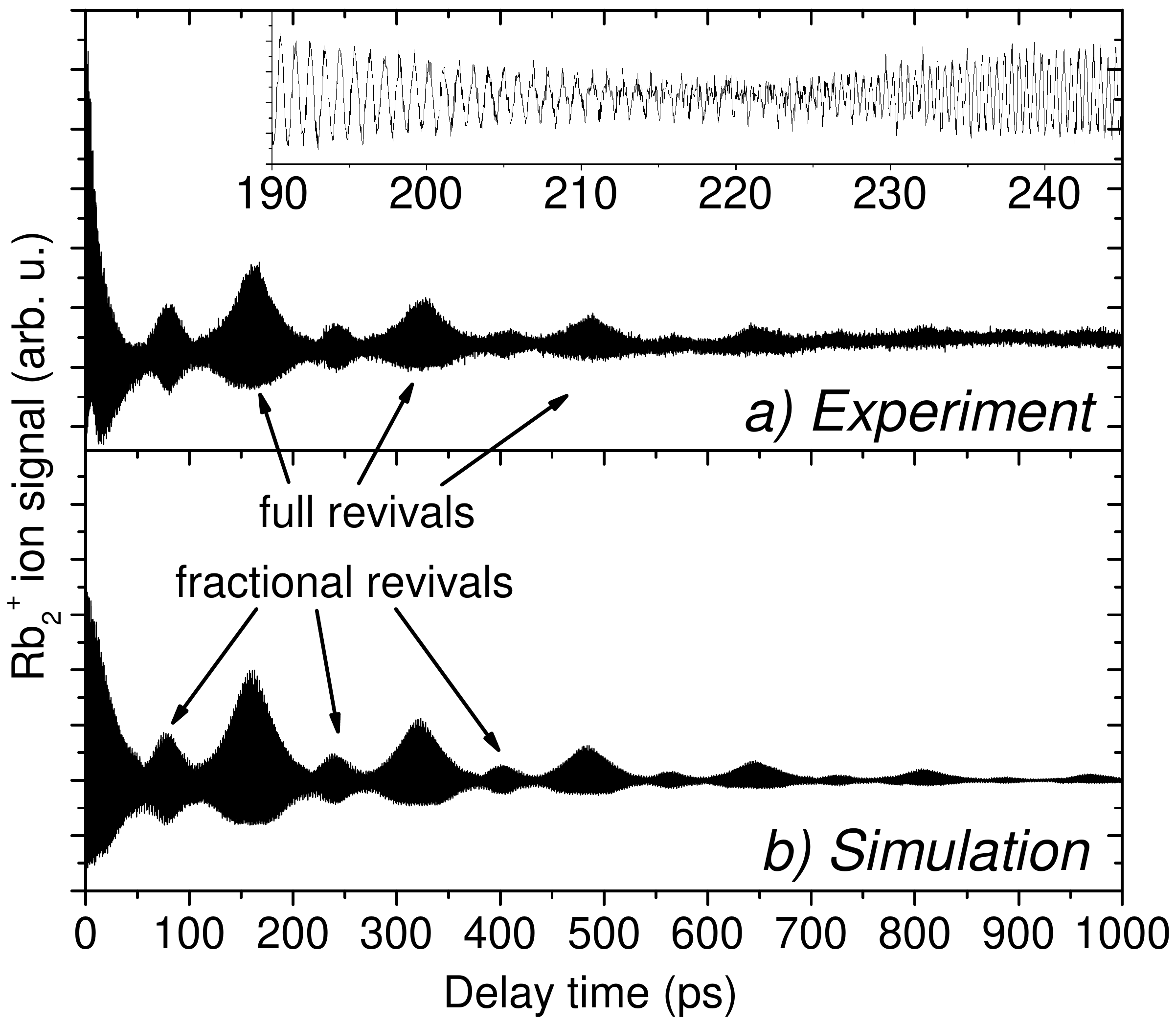}
\caption{Experimental (a) and simulated (b) pump-probe transients sowing vibrational wave packet propagation of Rb$_2$ formed on He nanodroplets recorded at $\lambda =1006$ nm~\cite{Gruner:2011}.}
\label{fig:Rb2}
\end{figure}
For Rb$_2$ in the lowest $1^3\Sigma_u^+$ and $1^3\Sigma_g^+$ triplet states, we were able to measure vibrational wave packet oscillations with high contrast over pump-probe delay times up to 1.5~ns~\cite{Mudrich:2009,Gruner:2011}. As for K$_2$, excellent agreement with quantum dynamics simulations was achieved under the assumption that dissipative coupling induces vibrational relaxation and dephasing of the coherent vibrational wave packets, see Fig.~\ref{fig:Rb2}. However, much weaker damping was observed (damping rate constant $\gamma\approx 0.5$ ns$^{-1}$) than for K$_2$, but comparable to Rb$^*$He. This may be related to the different spin-states of Rb$_2$ and K$_2$~\cite{Bovino:2009}, but also the atomic masses, the vibrational levels and oscillation frequencies were different. Moreover, the role of desorption of the dimers off the He droplets remains largely unresolved and requires further experimental investigations, e.~g. using ion and electron imaging.

\begin{figure}
\centering
\includegraphics[width=0.5\textwidth]{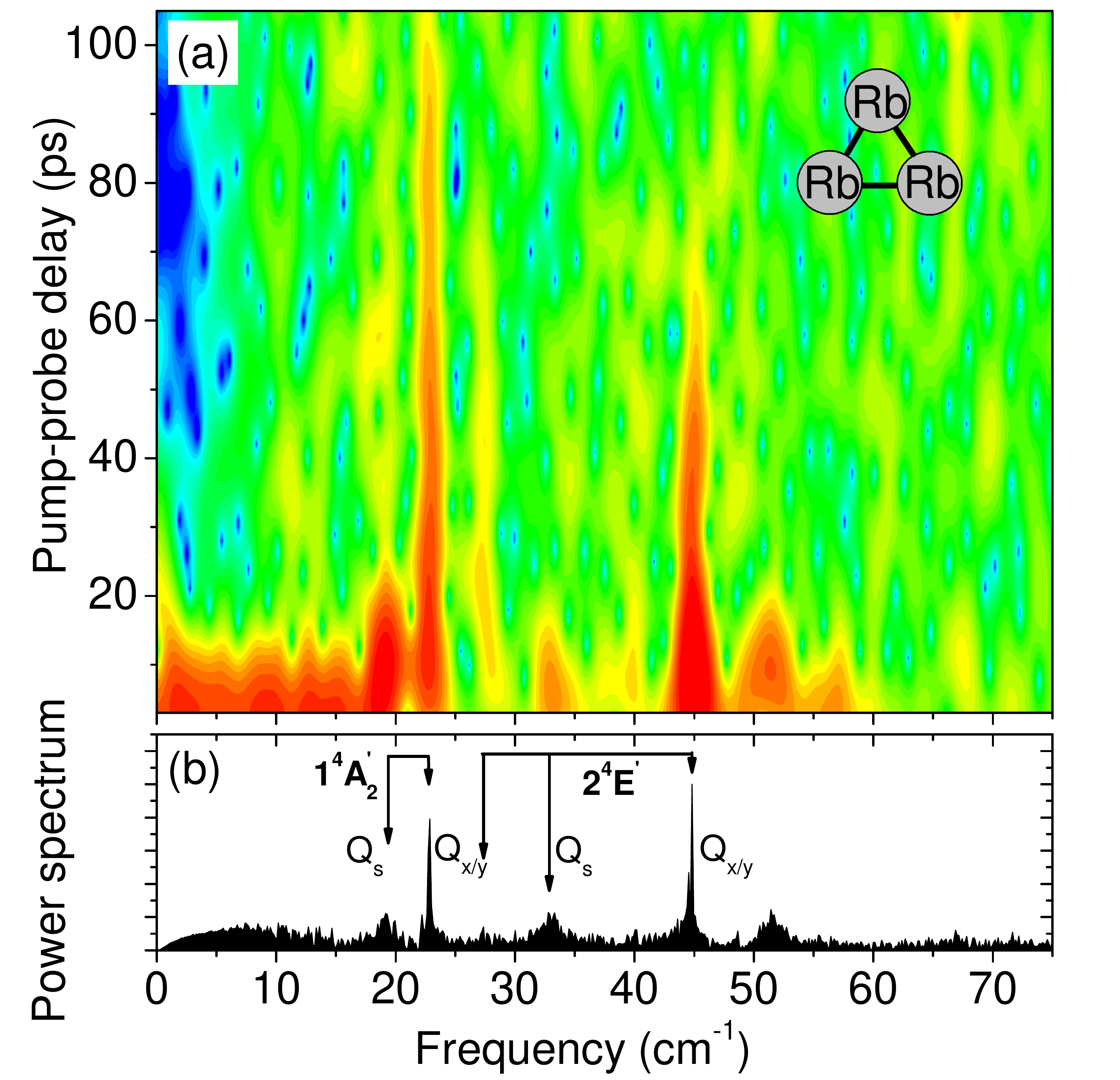}
\caption{(a) Sliding window Fourier analysis of the transient photoionization spectrum of Rb$_3$ trimers in quartet states recorded at $\lambda = 850$ nm. (b) Integral power spectrum. Adapted from~\cite{Giese:2011}.}
\label{fig:Rb3}
\end{figure}
Vibrational wave packet dynamics has also been measured for the Ak trimers K$_3$, Rb$_3$, and mixed species, formed on He nanodroplets~\cite{Giese:2011}. The individual vibrational frequency components are damped on a time scale $\sim 10$-$50$ ps, as can be seen in the sliding-window Fourier spectrum shown in Fig.~\ref{fig:Rb3} (a). clearly indicating effective coupling to the He bath. The interpretation of vibrational beat spectra of Ak trimers turned out to be much more involved than for the Ak dimers due to the complex vibronic structure of the heavy Ak trimers which are perturbed by Jahn-Teller and spin-orbit-couplings~\cite{Hauser:2011}. Nevertheless, all prominent frequency components measured for K$_3$ and Rb$_3$ could be assigned to the normal vibrational modes of the ground and first excited quartet states by comparing with high-level ab initio calculations carried out by A. Hauser in Graz (Fig.~\ref{fig:Rb3} (b))~\cite{Giese:2011}.

\begin{figure}
\centering
\includegraphics[width=0.5\textwidth]{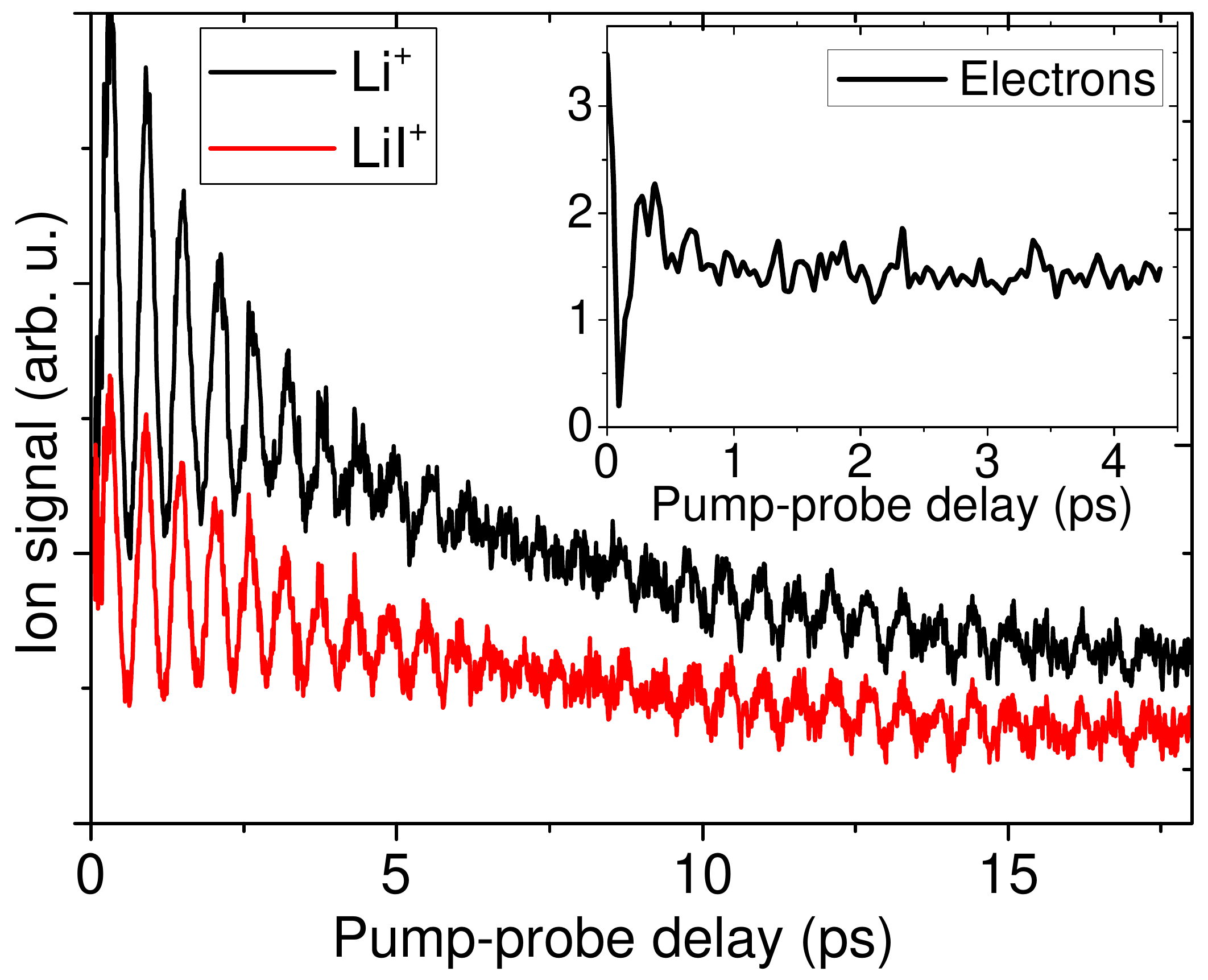}
\caption{Yield of Li$^+$ and LiI$^+$ ions created by femtosecond pump-probe ionization of LiI in an effusive beam at $\lambda =275$ nm. The inset shows the yield of photoelectrons from LiI prepared in He nanodroplets.}
\label{fig:LiI}
\end{figure}
Recently, we have tried to extend the scope of molecular systems to diatomics which are immersed inside He droplets so as to probe the effect of stronger system-bath couplings than for the surface-bound Ak molecules. Unfortunately, using the salt molecules NaI and LiI in He nanodroplets we have not succeeded in resolving vibrational wave packet dynamics, although wave packet oscillations are clearly visible for free molecules in the gas phase, as shown in Fig.~\ref{fig:LiI}. The yield of LiI$^+$ and Li$^+$ fragment ions, which oscillates due to the coherent vibrational wave packet motion in the excited $A$-state, falls off within $\sim 10$ ps due to predissociation~\cite{Schaefer:1986}. However, for LiI doped in He nanodroplets, only weak ion and electron count rates are measured which show no oscillations (inset of Fig.~\ref{fig:LiI}). We interpret this disappointing result by ultrafast vibrational relaxation induced by the He environment, which causes damping of the vibrational coherence on the time scale of the laser pulse duration ($\sim 150$ fs) and shifts transition frequencies out of the laser profile. Our finding seems to put severe limitations to the applicability of fs REMPI to surface-bound molecular species only. However, we speculate that larger organic molecules, for which the change in electronic and molecular structure upon electronic excitation is weaker and therefore vibronic spectra are less perturbed~\cite{Pentlehner:2010}, may still exhibit vibronic dynamics on time scales accessible to fs spectroscopy, similar to  molecules embedded in heavier rare gas matrices~\cite{Guhr:2007}.

Apart from these studies by our group, only Mg-doped He nanodroplets have been studied by fs spectroscopy in the group of K.-H. Meiwes-Broer, J. Tiggesb\"aumker in Rostock~\cite{Przystawik:2008,Goede:2013}. Based on linear absorption spectra and on fs REMPI transients of multiply doped He droplets, it was concluded that Mg atoms aggregate in He nanodroplets in an unusual way to form a foam-like structure where the metal atoms arrange themselves in a regular 10~\AA-spaced network separated by He atoms. This structure, which features a weakly shifted absorption line with respect to the Mg monomer in He droplets, is found to collapse upon electronic excitation to form metallic clusters. Thus, the transient mass spectra reveal a sharp drop of the yield of Mg$^+$ and small Mg$_n^+$ cluster ions within $\tau =350$~fs due to the decreased ionization cross section of Mg as the electronic properties evolve from atomic to
bulk-like~\cite{Przystawik:2008,Goede:2013}. Subsequent slow recovery of the Mg ion signals within $\sim 50$~ps was associated with the escape dynamics out of the He droplets.

Very recently, the group of H. Stapelfeldt in Aarhus has started to study the rotational dynamics of molecules embedded in He nanodroplets initiated by fs or nanosecond laser pulses~\cite{PentlehnerPRL:2013,Pentlehner:2013}. Contrary to the naive expectation that impulsively induced rotational coherences should be weakly damped based on the sharp lines in conventional IR spectra, the rotational dynamics was found to be significantly slowed down and rotational recurrences were completely absent~\cite{PentlehnerPRL:2013}. This indicates that transient system-bath interactions take place during the laser pulse and, possibly, correlations between the molecule and the He droplet are influencing the dynamics. For the case of adiabatic alignment induced by a weaker nanosecond IR pulse, however, nearly the same degrees of alignment were achieved for molecules in He droplets as compared to free molecules in the gas phase~\cite{Pentlehner:2013}. The authors point out the possibility of extending this approach to molecules that are tightly aligned under field-free conditions for long times by non-adiabatically switching off of the alignment pulse and by exploiting the slowed rotational dynamics in He droplets.

\section{\label{sec:EUV}EUV photoionization}
He has the highest ionization energy of all species ($E_i\approx 24.6$ eV). Therefore the direct PI of He nanodroplets requires radiation in the extreme ultraviolet (EUV) spectral range, which is traditionally provided by synchrotrons. More recent alternative sources of EUV radiation are the generation of high order harmonics using intense ultrashort near-infrared laser pulses as well as free-electron lasers (FEL). A complementary approach to photo-excitation or ionization, which is easier to realize experimentally but suffers from greatly reduced spectral resolution, is electron-impact ionization.
In this section we summarize the work on EUV PI of pure and doped He nanodroplets and we highlight recent developments.

\begin{figure}
\centering
\includegraphics[width=0.5\textwidth]{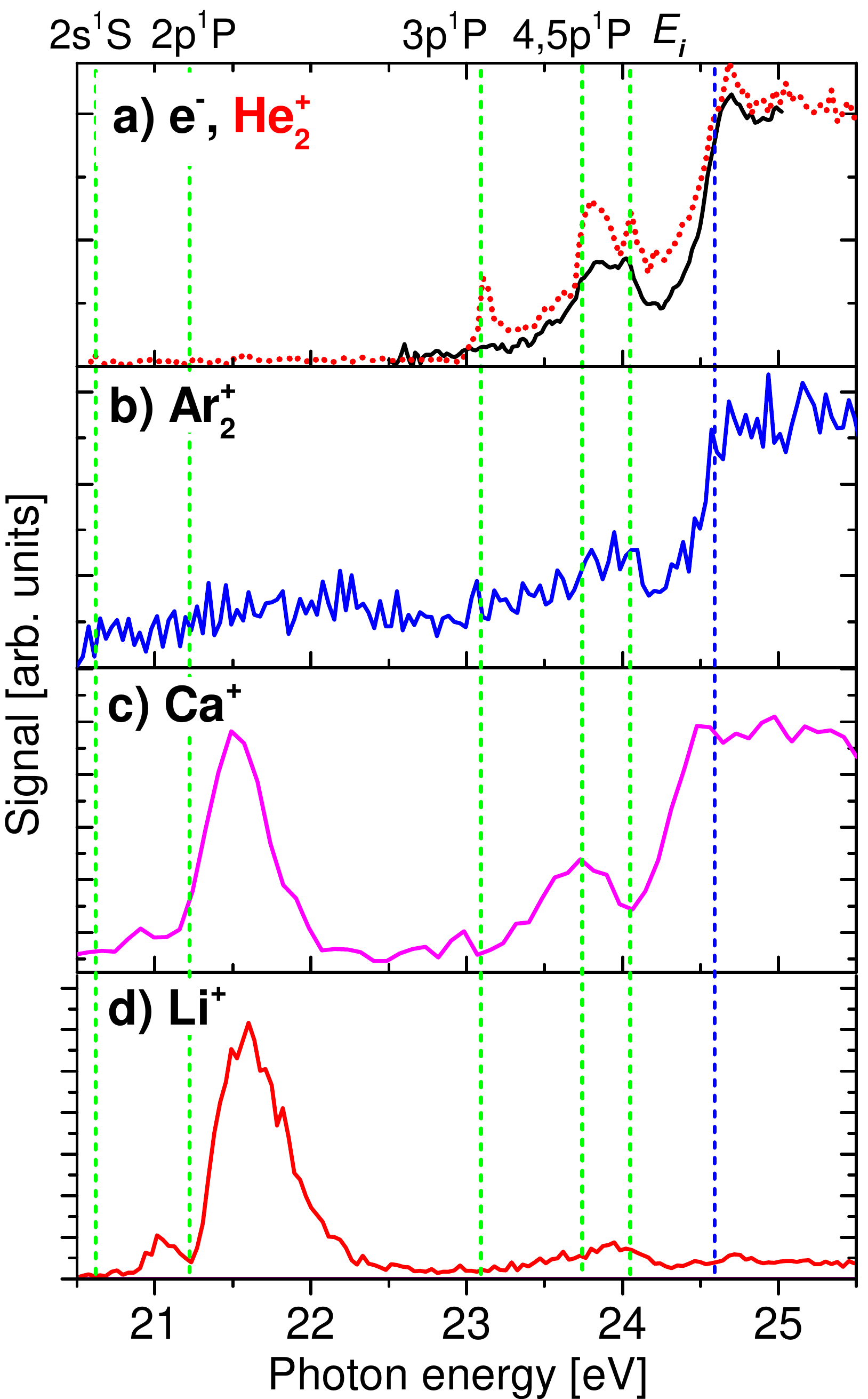}
\caption{Photon energy dependence of the yield of photoelectrons
and He$_2^+$ ions (a), Ar$_2^+$ (b), Ca$^+$ (c), and Li$^+$ ions (d) measured by illuminating
pure (a) and doped He nanodroplets (b-d). The vertical dashed lines indicate He atomic level energies. Adapted from~\cite{Buchta:2013}.}
\label{fig:Elettra}
\end{figure}
A first seminal study was reported by J. P. Toennies and coworkers in Berlin who used synchrotron radiation at photon energies $h\nu\sim E_i$ of He to assess the energetics and dynamics of photoexcitation and ionized He droplets~\cite{Froechtenicht:1996}. The key aspects of He droplet PI were already established in that work: Ionization occurs not only by a direct process at photon energies above the atomic $E_i$ but also by autoionization at photon energies below the atomic threshold in the range $23\lesssim h\nu\lesssim 24.6$ eV, see Fig.~\ref{fig:Elettra} (a). The latter ionization mechanism proceeds via the electronically excited states of the neutral droplet, which can be viewed as strongly perturbed atomic He states and therefore sensitively depends on the droplet size. The dominant ionization products in this regime are He$_2^+$ ions and small He$_n^+$ clusters as well as large cationic clusters with $n>10^3$. The decay by fluorescence emission is more probable than by ionization following the photoexcitation process. In droplets with embedded SF$_6$ molecules, the dopants are ionized indirectly by Penning-like excitation transfer ionization via He$^*$ ``excitons'' which leads to a large ion signal on the mass of the embedded species whereas no evidence for direct PI of dopants was found.

These studies have been refined in a series of synchrotron experiments carried out by D. Neumark and coworkers in Berkeley. By applying photoion and electron imaging detection, detailed insights into the relaxation processes following photoexcitation or ionization of He droplets were obtained. Most strikingly, the photoelectron spectra recorded below the He atomic $E_i$ are dominated by very low energy electrons, with electron kinetic energies $<0.6$ meV~\cite{Peterka:2003,Peterka:2007}. The exact formation mechanism of these ZEKE electrons is currently under discussion. At that time the occurrence of ZEKEs was interpreted to be due to vibrational autoionization of He$_N^*$ Rydberg states. A more detailed discussion based on further studies at higher photon energies also included the formation and decay of an electron bubble~\cite{Peterka:2007}. More recent experiments confirmed the appearance of ZEKEs as a general phenomenon when releasing electrons in He droplets, even when ionizing dopants with low $E_i$ at conditions where the electronic excitation of He is excluded~\cite{Fechner:2012}. Obviously, part of the emitted electrons experience efficient inelastic scattering to dissipate their entire kinetic energy. Since we have observed ZEKEs also in larger water clusters \cite{LaForge:2014a}, the mechanism is most likely not related to the electronic structure of He droplets nor to electron bubble formation or superfluid properties. Calculations modeling the electron dynamics show that the equilibration of electrons by scattering processes transfers a large fraction of electrons in Rydberg-type orbits where due to the low density, interaction with the rest of the system becomes unlikely \cite{Fennel:2014,Saalmann:2014}.

Photoelectron spectra of He droplets measured above the He atomic $E_i$ featured a fraction of electrons with a kinetic energy up to $\lesssim 0.5$ eV higher than that from free He atoms. This observation was rationalized by the direct ionization into bound states of the He$_2^+$ molecular cations. Our recent ion mass-resolved photoelectron imaging measurements support this interpretation~\cite{BuchtaJCP:2013}. It implies that at least part of the charges immediately localize as He$_2^+$ instead of freely migrating as He$^+$ ``holes'' through the droplet due to resonant charge hopping~\cite{Halberstadt:1998}.

Photoelectron spectra of He nanodroplets doped with rare-gas atoms ionized at photon energies below the He atomic $E_i$ have revealed that Penning ionization of the dopants proceeds either by direct excitation transfer from the photo-excited 1s2p-state of He or from the lower-lying, long-lived 1s2s-state which is populated by droplet-induced electronic relaxation~\cite{Wang:2008}. The resulting electrons were found to undergo considerable electron-He scattering so as to lose $\sim 10$ eV of kinetic energy. For large droplets, a gap in the photoelectron spectrum appeared at energies $\lesssim 1$ eV, which was interpreted as indication for the development of the droplet analogue of the conduction band in bulk liquid He. In contrast, the photoelectron spectra correlated to Penning ionization of SF$_6$ showed only minor differences when compared to those of free SF$_6$, pointing at optical-like electronic dipole interaction to be active rather than ``traditional'' Penning ionization as in collisions involving metastable atoms~\cite{Peterka:2006}. Ion-imaging of the Penning ionization products revealed considerably slower velocity distributions of the escaping SF$_5^+$ fragment ions as compared to the gas-phase, in agreement with a binary collision model.

In recent experiments at Elettra Synchrotrone, Trieste, we have extended the previous synchrotron work by implementing photoelectron-photoion coincidence detection. In this way, photoelectron spectra and angular distributions were measured in coincidence with specific ion masses~\cite{Buchta:2013,BuchtaJCP:2013}. Argon (Ar) dopants which are immersed in the droplet interior are predominantly ionized by charge transfer occurring upon ionization of the He droplet in the reaction 
\[
\mathrm{He}^+ + \mathrm{Rg}\rightarrow \mathrm{He} + \mathrm{Rg}^+ + \Delta E.
\]
Accordingly, the measured yield of Ar$_2^+$ dopant ions as a function of photon energy closely follows the yield of He ions, see Fig.~\ref{fig:Elettra} (b). The most surprising finding is that Ak metal dopants, which are weakly bound at the surface of He nanodroplets, are efficiently Penning ionized upon excitation of the lowest excited states of the host droplets at $h\nu =21$ and $21.5$ eV, see Fig.~\ref{fig:Elettra} (d). This indicates rapid migration of He$^*$ excitations to the droplet surface, followed by electronic relaxation, and eventually energy transfer to the Ak dopants in a process of the type~\cite{Buchta:2013}
\[
\mathrm{He}^* + \mathrm{Ak}\rightarrow \mathrm{He} + \mathrm{Ak}^+ + e^-.
\]

Alkaline-earth metals are an intermediate case in the sense that earth-alkaline atoms such as Ca reside inside the He droplet surface layer~\cite{Hernando:2008}. Correspondingly, the ionization via Penning and charge transfer reactions proceeds with similar probabilities (Fig.~\ref{fig:Elettra} (c)). Photoelectron spectra measured in coincidence with dopant ions indicated that Penning ionization occurs predominantly out of relaxed electronic states of He$^*$, similarly to the previous observations~\cite{Wang:2008}.

The advent of laser-based fs EUV light sources has made it possible to perform table-top experiments using ultrashort EUV pulses. Using a EUV-NIR pump-probe setup in combination with electron and ion imaging detection, the  group of O. Gessner in Berkeley has succeeded in performing time-resolved measurements on pure He droplets~\cite{Kornilov:2010,Kornilov:2011,Buenermann:2012,BuenermannJCP:2012}. These measurements have revealed a complex relaxation dynamics to be initiated by the pulsed EUV excitation into perturbed Rydberg states of the droplets (``bands''). Subsequent relaxation proceeds via various channels, involving both inter- and intra-band relaxation into lower lying droplet states. Based on the time-resolved electron and ion kinetic energy distributions, intra-band relaxation was rationalized by the expulsion of localized excited He$^*$ Rydberg atoms out of the droplets. Specifically, He atoms in orbitally aligned 1s4p-states and in unaligned 1s3d states were found to be the dominant fragments after exciting large droplets at a photon energy of $23.6\pm 0.2$ eV. The ejection time scales of atoms in 1s4p-states and 1s3d-states were determined to be $<120$ fs and $\sim 220$ fs, respectively. The very low energy electron component was observed with a rise time of 2-3 ps~\cite{Kornilov:2011}.

Based on the mentioned PI experiments as well as on dispersed fluorescence emission measurements~\cite{Joppien:1993,Haeften:1997,Haeften:2001,Haeften:2005,Haeften:2011}, the photoexcitation or ionization dynamics of He nanodroplet can be classified into the following three regimes:

(i) At photon energies $20.5<h\nu<23\,$eV, He nanodroplets are excited with large cross sections into perturbed excited states derived from the 1s2s$^1$S and 1s2p$^1$P He atomic levels. Fast droplet-induced intra-band and inter-band relaxation as well as He$_2^*$ excimer formation follows the excitation~\cite{Haeften:1997,Haeften:2005,Kornilov:2011,Buenermann:2012}. Due to the repulsive interaction between excited He$^*$ or He$_2^*$ and the He environment the excitation migrates to the surface presumably involving both resonant hopping of the electronic excitation as well as nuclear motion of the excited He$^*$ atom~\cite{Scheidemann:1993,Buenermann:2012,BuenermannJCP:2012,Buchta:2013}. Depending on the size of the He droplet, the He$^*$(1s2p$^1$P)-state is trapped at the surface and eventually relaxes into the long-lived 1s2s$^{1,3}$S-states or into vibrationally excited He$_2^*$ molecules~\cite{Buchenau:1991}. The latter are subject to vibrational relaxation by coupling to the He droplet and eventually evaporate off the droplet surface. Since the photon energy falls below the ground state energy of He$_2^+$ and larger cationic He clusters, the excited He droplets cannot decay by autoionization. Fluorescence emission is the only decay channel for pure He droplets. In doped He droplets excited into the 1s2s$^1$S and 1s2p$^1$P droplet states, dopants can be indirectly ionized by a Penning ionization process. While this process is extremely efficient for Ak metal dopants which are bound at the droplet surface, it is much less efficient for rare gas atoms which are immersed in the droplet interior~\cite{Froechtenicht:1996,Buchta:2013,Kim:2006,Wang:2008}.

(ii) At photon energies $23<h\nu<24.6\,$eV, the droplet response is even more complex. In addition to the aforementioned relaxation channels, the emission of He$^*$ and He$_2^*$ in Rydberg states dominates~\cite{Haeften:2005,Buenermann:2012,BuenermannJCP:2012}, while the fraction of He$_2^*$ dimers increases with rising excitation energies~\cite{Haeften:1997,Haeften:2005}. At $h\nu>24\,$eV the population of triplet states of He was also observed presumably due to electron-ion recombination~\cite{Haeften:1997,Kornilov:2010}. As a further relaxation channel, autoionization of He droplets opens up as a nonradiative decay channel at $h\nu>23\,$eV which competes with fluorescence emission. Both small ionic fragments (He$_n^+$, $n\leq 17$) as well as large cluster ions ($N\gtrsim 10^3$) are formed by autoionization~\cite{Froechtenicht:1996}. A peculiarity of the ionization of He droplets below the He atomic $E_i$ is the emission of electrons with very low kinetic energy $<1\,$meV as seen in photoelectron imaging experiments~\cite{Peterka:2003,Peterka:2007}. Recent time-resolved photoelectron and photoion imaging experiments have revealed the dynamics of various relaxation processes in this regime~\cite{Kornilov:2010,Kornilov:2011,Buenermann:2012,BuenermannJCP:2012}. In this regime, dopant ionization can proceed both by excitation transfer (Penning ionization) or by charge transfer following droplet autoionization~\cite{Froechtenicht:1996,Buchta:2013,Kim:2006}.

(iii) At photon energies $h\nu>24.6\,$eV, that is above $E_{i,\mathrm{He}}$, He$^+$ ions (positive holes) are created in the droplets. The positive charges subsequently migrate through the He droplet by resonant hopping and eventually localize by forming He$_2^+$ molecular ions or by ionizing a dopant if present~\cite{Stace:1988,Scheidemann:1993,Halberstadt:1998,Seong:1998}. The internal energy of the newly formed ion as well as the binding energy liberated upon formation of snowball structures (He atoms tightly bound around the ion core) is believed to stop the charge-hopping process and to cause massive droplet fragmentation. Therefore, He$^+$ largely from background He atoms and He$_2^+$ from droplets are the dominant species appearing in the mass spectra~\cite{Froechtenicht:1996,Peterka:2007,Kim:2006,Buchta:2013,BuchtaJCP:2013}.
%Dopants are efficiently ionized indirectly by charge transfer.

Recently, we have reported the first study of He droplet ionization using intense EUV radiation generated by a free-electron laser~\cite{LaForge:2014,Ovcharenko:2014}. This study probes the transition from weakly to multiply excited He droplets which evolve into a palsma-like state by many-body interactions. The work connects to the topic of strong-field ionization of clusters creating highly ionized, confined states of matter (nanoplasma), which we discuss in more detail in the following section.

\section{\label{sec:StrongField}Strong-field ionization}
He nanodroplets are mostly used for isolating molecules at low temperature in a transparent and extremely inert environment. The particularly favorable properties of He droplets originate in the extremely high excitation and ionization energy of He in combination with the extremely low droplet temperature and the resulting superfluid state. In a certain sense, He droplets can be viewed as nano-cryostats which efficiently cool embedded molecules without greatly perturbing their spectra, or even as nano-cryo reactors which can be used for synthesizing specific molecular complexes and for studying chemical reactivity in a weakly influencing environment.

However, the opposite behavior occurs when He droplets doped with a few heavier rare-gas atoms are illuminated by intense near-infrared light~\cite{MikaberidzePRL:2009}. Initiated by tunnel-ionization of the dopant atoms which act as seeds, a He nanodroplet is turned into a nanoplasma, a highly reactive medium, which greatly enhances the ionization state, the fragmentation rate and ion kinetic energies of the dopant cluster.

Strong-field ionization of heavier rare-gas clusters has been thoroughly investigated both experimentally and theoretically for over two decades now~\cite{Fennel:2010,Saalmann:2006}. Most frequently, near-infrared fs laser pulses with a central wavelength of $\sim 800$ nm (photon energy $\sim 1.5$ eV) from Ti:Sapphire lasers have been employed. These pulses are said to be ``intense'' ($10^{13}$ - $10^{16}$ Wcm$^{-2}$) when the peak electric field is comparable to the Coulomb field binding electrons in an atom~\cite{Krainov:2002}.

Under the influence of such an external electric field, bound electrons can tunnel out through the potential barrier which results from the sum of the native atomic potential and that due to the laser pulse. The laser field then acts as a strong time-varying external driving field which forces the released electrons into oscillatory motion through the cluster. The resulting electron-impact ionization in the field of the created ionic cores enhances the buildup of a confined plasma-like state with a charge density $\rho$. During this laser-driven ionization process, a large fraction of electrons released from their parent atoms or ions remain trapped in the space charge potential of the cluster (inner ionization). The critical phase of light-matter interaction sets in as the plasma expands and the dipolar eigenfrequency of the plasma (``plasmon''), which for the spherical case is given by
\[
\omega_P=\sqrt{\frac{e\rho}{3\varepsilon_0 m_e}},
\]
meets the frequency of the driving laser field $\omega_L$~\cite{Fennel:2010,Saalmann:2006}. Here $e$ and $m_e$ denote the electron charge and mass, respectively, and $\varepsilon_0$ is the vacuum permittivity. Under such resonance conditions ($\omega_P=\omega_L$), the nanoplasma becomes highly light absorbing, by far more than atomic jets or planar solid targets~\cite{Ditmire:1997,Zweiback:1999}. Consequently, the nanoplasma heats up dramatically and emits electrons and ions (outer ionization). Very high ion charge states~\cite{Snyder:1996,Lezius:1998,Prigent:2008}, electron energies up to multi-keV~\cite{Shao:1996,Springate:2003,Kumarappan:2003,Namba:2006} and ion kinetic energies up to MeV~\cite{Lezius:1998,Springate:2000,Kumarappan:2001}, and even EUV and x-ray photons have been detected~\cite{McPherson:1994,Parra:2000,KumarappanPRA:2001,Prigent:2008}.

Composite clusters containing a heavy rare-gas and a second molecular component have been studied by the group of M.~Krishnamurthy in Mumbai, India. For the cases of doping argon clusters with H$_2$O or CS$_2$ molecules a significant enhancement of the yield of highly energetic electrons and ions as well as of characteristic X-ray emission was observed~\cite{Jha:2005,Jha:2009}. This enhancement of the plasma variables was attributed to an increase of the electron density earlier in the laser pulse as compared to pure argon clusters resulting in more efficient energy absorption and enhanced nanoplasma heating~\cite{Jha:2008,Jha:2009}.

Strong-field ionization of doped He nanodroplets has not been studied until the year 2001, when K.-H. Meiwes-Broer, J. Tiggesb\"aumker and coworkers in Rostock started to extend their work on metal clusters to clusters aggregated in He droplets~\cite{Diederich:2001,Doeppner:2001}. On the one hand, producing He droplet beams requires substantially more technical effort (cryogenic He source, large pumps in the case of a continuous beam) than producing clusters out of the heavier rare gases, which can be generated by conventional pulsed nozzles around room temperature. Traditionally, those groups focusing on the development of ultrashort laser techniques tend to be repelled from involved molecular beam machinery. On the other hand, pure He droplet are less attractive targets for strong-field ionization studies due to the high threshold intensity needed for singly ionizing He ($1.5\times 10^{15}$ Wcm$^{-2}$~\cite{Augst:1989}) and the low number of only two electrons which each He atom can at most contribute to building up a nanoplasma. However, He droplets doped with heavier species have recently revealed quite diverse strong-field ionization dynamics related to the extremely large differences in ionization energies of dopants and He matrix and to the controlled localization of dopants inside or at the surface of the droplets. This will be discussed in more detail below.

\begin{figure}
\centering
\includegraphics[width=0.7\textwidth]{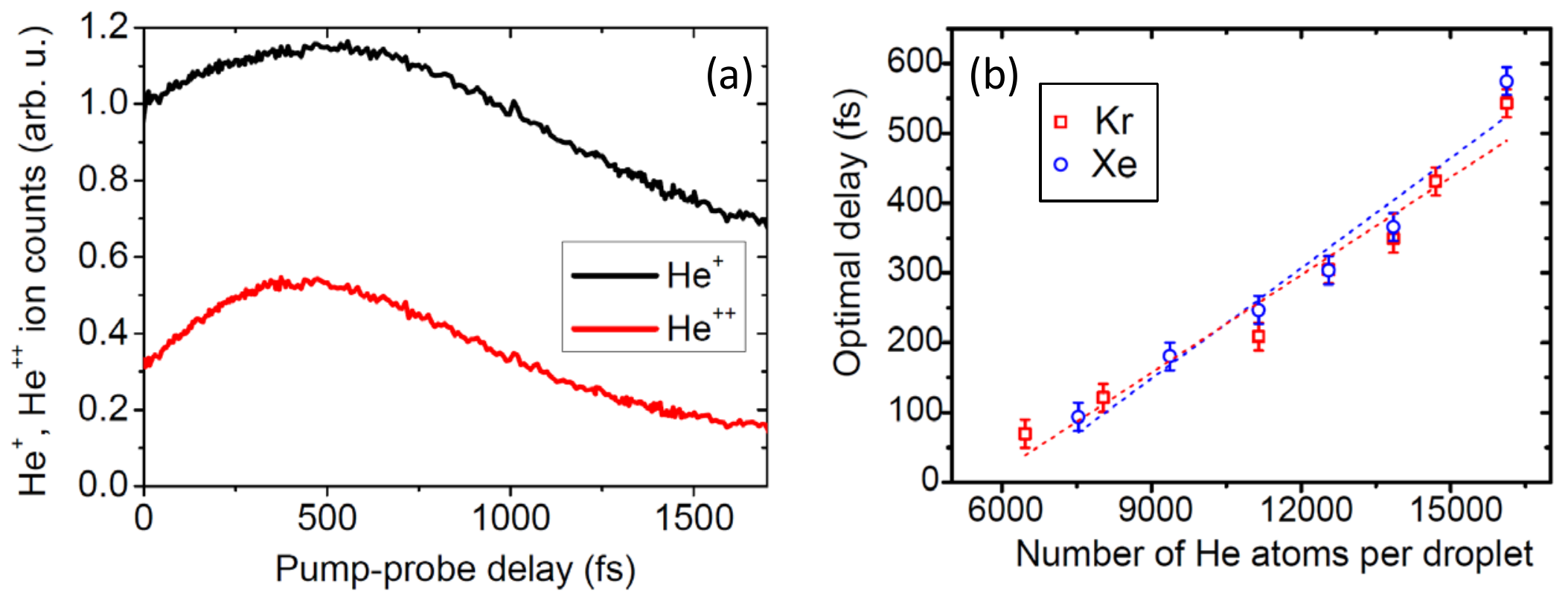}
\caption{(a) He$^+$ and He$^{2+}$ ion yields as a function of pump–probe delay when He
nanodroplets doped with Xe atoms are exposed to two identical pulses ($\sim 10$ fs)
of peak intensity $7\times 10^{14}$ Wcm$^{-2}$. The mean droplet size is 15\,000 He atoms,
and the mean number of Xe dopants is $15\pm 3$. (b) Optimal delay times for doped He nanodroplets of various sizes. Adapted from~\cite{Krishnan:2012}}
\label{fig:NanoplasmaPP}
\end{figure}
Since the focus of the Rostock group has been on strong-field ionization dynamics of metal clusters, He nanodroplets have mostly been regarded as an alternative method for generating metal clusters~\cite{Doeppner:2005,Tiggesbaumker:2007}. However, experiments with single intense laser pulses of variable duration and with dual pulses applied to He droplets containing lead (Pb) and Ag clusters revealed a significant influence of the He environment on the ionization dynamics of the metal core~\cite{Doeppner:2003,Doeppner:2007}. (i) The condition for resonant charging is
reached earlier in time, and (ii) the maximum recorded charge state of the metal ions is reduced. Observation (i) was rationalized by more efficient initial non-resonant charging of the metal kernel in the presence of the He environment due to the presence of additional electrons generated by electron-impact ionization of He surrounding the metal core nanoplasma. This leads to a faster expansion and thus to an earlier resonance between the photon and the plasmon energies. The observed reduction of the maximum charge state (ii) was explained by charge transfer from neutral or singly ionized He atoms into free energy levels of the highly charged metal ions. As a further consequence of the metal-He interaction, in large droplets caging of ionization fragments was observed, which induces the reaggregation of the metal (M) clusters~\cite{DoeppnerJCP:2007,Doeppner:2007}. The polarization interaction of metal ions with the surrounding He atoms causes the formation of MHe$_n^+$ snowball complexes~\cite{DoeppnerJCP:2007,Doeppner:2007,Tiggesbaumker:2007}, which have been discussed in section~\ref{sec:MassSpectra}.

The active role of the He shell in the strong-field ionization process of doped He nanodroplets was first pointed out by A. Mikaberidze, U. Saalmann and J.-M. Rost in Dresden~\cite{Mikaberidze:2008,MikaberidzePRL:2009}. Using classical molecular dynamics simulations, the authors studied the energy absorption of a two-component Xe$_{100}$He$_{1000}$ core-shell cluster illuminated by ultrashort dual pulses. According to their findings, two distinct resonance maxima appear in the absorption transient as a function of the pulse delay as a result of different speeds of expansion of the heavy Xe and the light He ionic components. Similar results were presented by C. Peltz and T. Fennel in Rostock who theoretically predict separate resonance features in the transient photoelectron spectra for a Xe$_{309}$He$_{10000}$ cluster but not for the transient ion spectra~\cite{Peltz:2011}. At optimal pulse delay the ionization state of the Xe dopants is higher in the presence of the He shell than for the free Xe cluster, mostly due to more efficient inner ionization by the first laser pulse.

Recent experiments confirm the model calculations in that doping the He droplets with heavier rare-gas atoms efficiently ignites a He nanoplasma and a plasmon resonance feature is clearly visible in the transient He$^+$ and He$^{2+}$ ion yields, see Fig.~\ref{fig:NanoplasmaPP} (a)~\cite{Krishnan:2012}. By using few-cycle pulses of a duration $\sim 10$ fs changes of the optical response during the pulses due to the ionic motion was excluded. The position of the maximum shifts from a delay time $\tau_{\mathrm{max}}\approx 100$ fs for small He nanodroplets He$_N$, $N\approx 6000$ to $\tau_{\mathrm{max}}\approx 500$ fs for $N\approx 15\,000$ (Fig.~\ref{fig:NanoplasmaPP} (b)), in good agreement with the numerical simulation.

\begin{figure}
\centering
\includegraphics[width=0.5\textwidth]{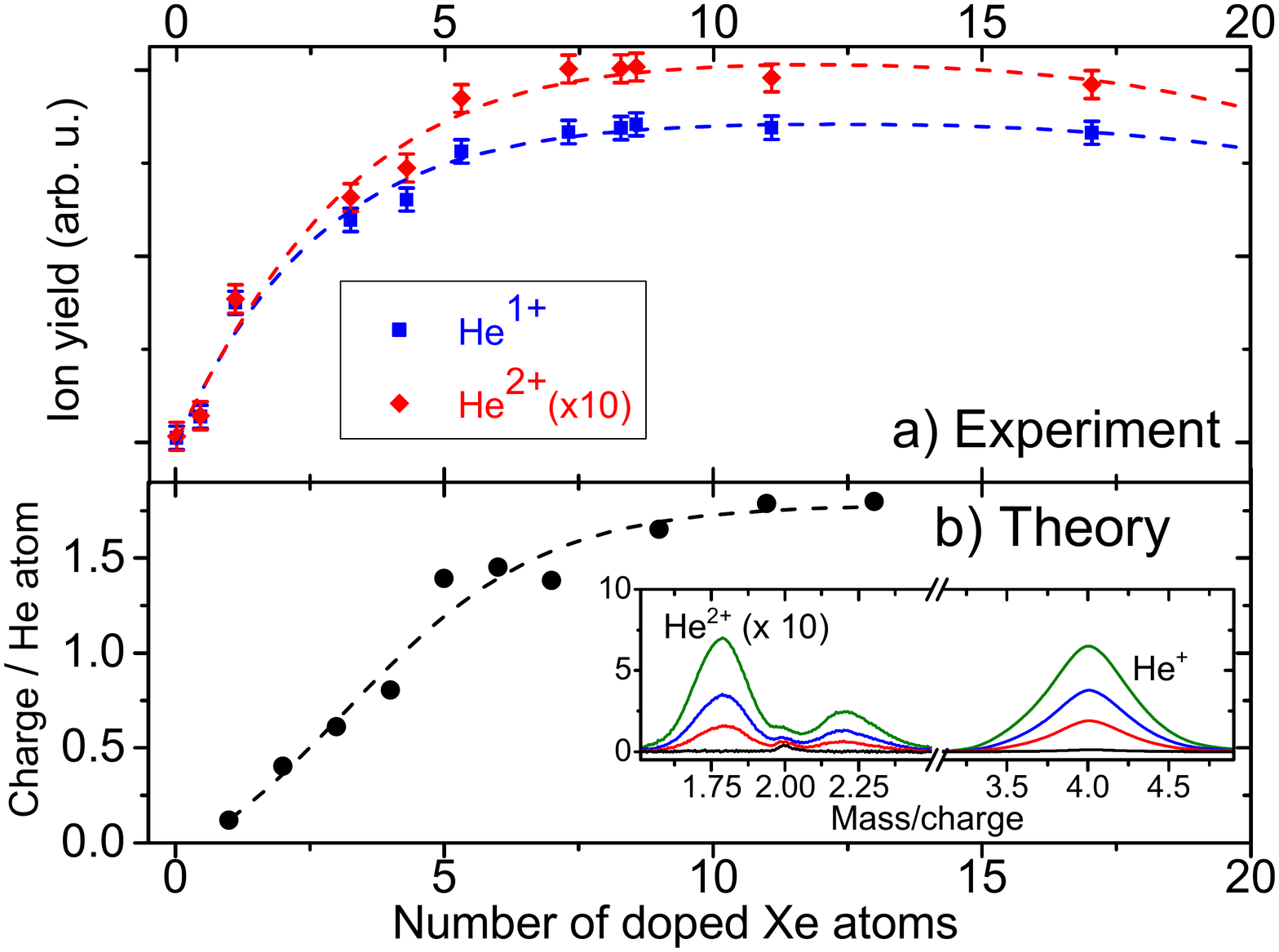}
\caption{(a) Yields of He$^+$ and He$^{2+}$ ions as a
function of the mean number of Xe dopants in a nanodroplet
containing 15\,000 He atoms at a peak laser intensity of
$7\times 10^{14}$ Wcm$^{-2}$. (b) Numerical calculation of the charge per
He atom as a function of the number of Xe atoms in a droplet
containing 4000 He atoms. All lines are to guide the eye. Inset:
Time-of-flight mass spectra of He$^+$ and He$_2^+$ ions for different
numbers of dopant Xe atoms (K) as shown in the legend~\cite{KrishnanPRL:2011}.}
\label{fig:Siva}
\end{figure}
A different type of energy absorption resonance was recently predicted for Xe-doped He nanodroplets by A. Mikaberidze, U. Saalmann and J.-M. Rost~\cite{MikaberidzePRL:2009} and experimentally demonstrated~\cite{KrishnanPRL:2011}. It was shown that only a few ($\lesssim 10$) Xe atoms suffice to spark the complete inner ionization of the He droplet at laser intensities below the ionization threshold of pure He droplets. Subsequent resonant absorption is enabled by an unusual cigarshaped nanoplasma which is elongated along the laser polarization axis within the droplet. Correspondingly, a nearly step-like increase of the yield of He ions is observed experimentally when doping the droplets with a few Xe atoms, in accordance with simulations (Fig.~\ref{fig:Siva}). Since this resonance phenomenon relies purely on the electron dynamics and not on the expansion of ionic cores, resonance conditions are reached within just a few optical cycles of the driving laser field. Therefore, the direct sampling of this resonance in a time-resolved experiment has not yet been possible.

\begin{figure}
\centering
\includegraphics[width=0.6\textwidth]{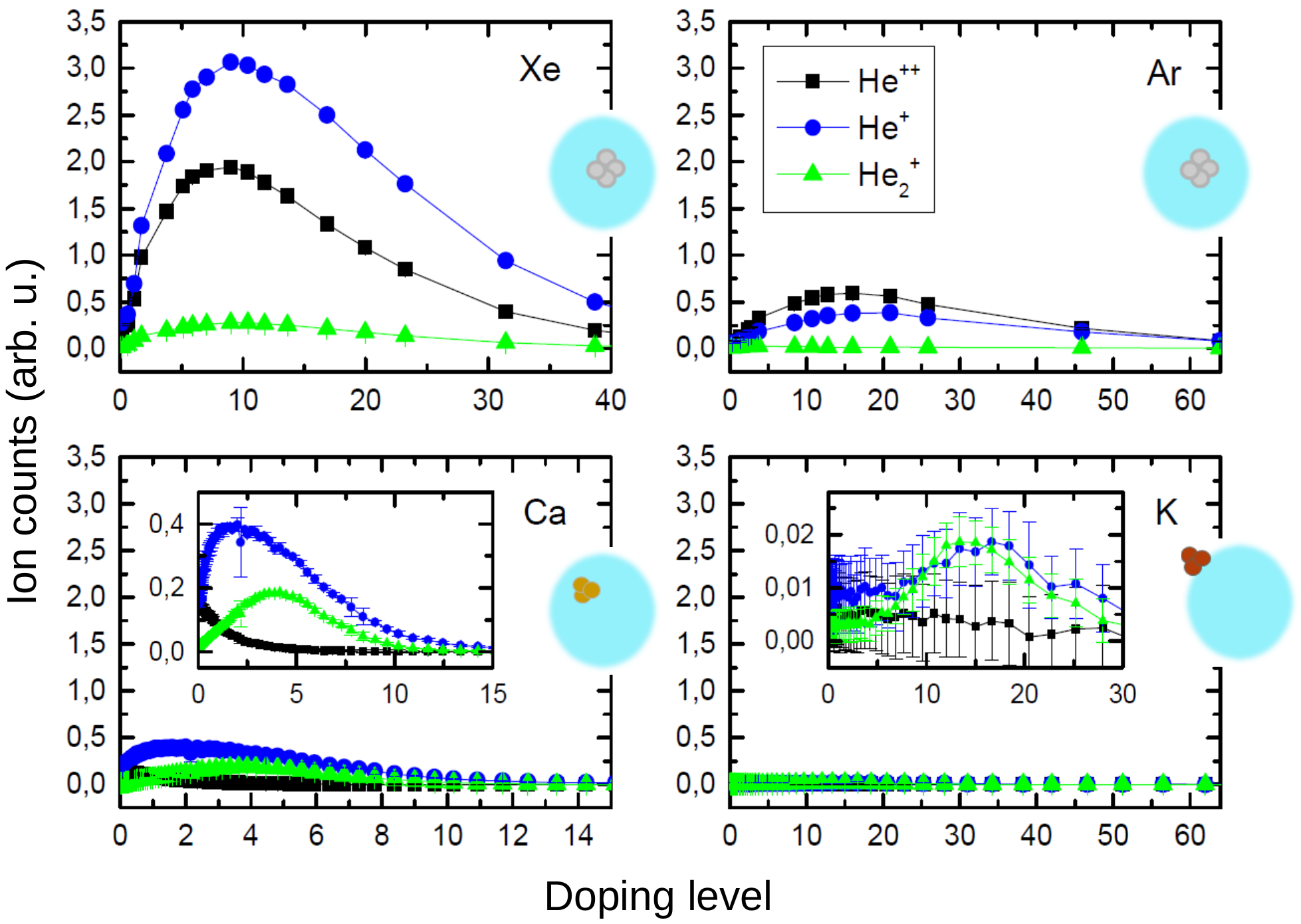}
\caption{Yields of He$^+$, He$^{2+}$, and He$_2^+$ ions as a
function of the mean number of dopants of various species as indicated in the legends in He nanodroplets containing 15\,000 He atoms at a peak laser intensity of
$7\times 10^{14}$ Wcm$^{-2}$.}
\label{fig:Species}
\end{figure}
In the experiments with two-component heavy rare-gas clusters the exact structure of the clusters is not very well defined~\cite{Jha:2009}. In contrast, He nanodroplets offer the possibility of studying strong-field nanoplasma ignition for various dopant species which are located in well-defined positions inside or at the surface of He droplets. Surprisingly, we observe drastically changing yields of ionization products when replacing Xe or Ar dopants (inside) by Ca (inside surface layer) or by K (on the surface), see Fig.~\ref{fig:Species}. Ongoing model calculations indicate that besides the position relative to the droplet surface, the differing ionization energies and the details of the doping and cluster aggregation process play a role~\cite{Heidenreich:2014}.

With the advent of intense EUV and X-ray radiation sources provided by free-electron lasers (FEL), the dynamics of nanoplasmas is being studied at increasingly high photon energies reaching up to the keV range. At such high photon energies, however, the physics of cluster ionization is rather different as compared to plasmon-enhanced charging of clusters illuminated by near-infrared pulses. In particular, multistep PI becomes the dominant absorption process and plasma heating has no significant effect~\cite{Bostedt:2008}. In recent experiments performed in a collaboration at the LDM beamline \cite{Lyamayev:2013} at the FEL FERMI@Elettra in Trieste, using intense tunable EUV radiation, the autoionization dynamics of multiply excited He nanodroplets was studied~\cite{LaForge:2014,Ovcharenko:2014}. Extremely high ionization rates measured upon excitation below the He ionization threshold as well as photoelectron spectra which are characteristic for thermal electron emission indicate that a novel many-body autoionization process is active which is related to inter-atomic Coulombic decay (ICD)~\cite{Kuleff:2010,Ovcharenko:2014}.

Due to the low ponderomotive energy of electrons in the EUV field of the order of meV this regime would be classified as the weak-field regime. Nevertheless, owing to the high intensities  available at FELs the majority of He atoms are electronically excited within tens of fss. The subsequent ultrafast collective autoionization of the whole cluster generates a highly ionized nanoplasma state similarly to the case of strong-field NIR ionization. Ongoing EUV experiments on doped clusters including fs time-resolve measurements let expect exciting insight into the details of such extreme states of matter.

%\begin{figure}
%\includegraphics[width=8cm]{Fig4_DifferentialArKrXe.pdf}
%\caption{\label{fig:differential} (Color online) (a) Experimental (symbols) and calculated (lines) differential cross sections of Ar, Kr, Xe, and SF$_6$ elastic backward scattering with trapped Li atoms as a function of relative velocity. (b) Integral cross sections for Li+Kr collisions.}
%\end{figure}

\section{\label{sec:Summary}Summary and prospects}
This article reviews recent achievements in the research of He nanodroplets with a focus on  photoionization techniques. In the past years, various schemes of photoionization in combination with sophisticated electron and ion detection methods have been developed, including resonance-enhanced ionization, photoelectron spectroscopy, and ion detection with mass, velocity and time-resolution. The introduction of electron and ion imaging as well as femtosecond techniques has given unprecedented insight into the dynamics of He droplets doped with various atomic and molecular species and subjected to laser excitation.

While the guest-host interaction of ground state atoms and molecules doped into He nanodroplets are generally weak, a complex photochemical dynamics is initiated by the electronic excitation or ionization of the dopants. In the simplest case, the full evaporation of the surrounding He by energy transfer from the excited molecules to the droplets provides a new type of action spectroscopy of cold molecules~\cite{Loginov:2008}. More surprisingly, molecular ions were found to literally pop out of the droplets upon vibronic excitation~\cite{Smolarek:2010}, which may be related to the formation of large cavitation bubbles upon pulsed energy deposition inside bulk superfluid He~\cite{Popov:2013}, but largely remains an unresolved puzzle. 

In contrast, the solvation and desolvation dynamics of metal atoms attached to He droplets is now well characterized even for excited levels approaching the ionization threshold and above~\cite{LoginovPRL:2011,Lackner:2011,Zhang:2012,Mateo:2013,Loginov:2014}. Surprisingly, the dopant-droplet complex can be quite accurately represented by a pseudo-diatomic molecule in many cases~\cite{Hernando:2012,Fechner:2012}. Only the observed fast droplet-induced relaxation of the excited metal atom into lower-lying states still remains and open issue~\cite{Loginov:2014,Vangerow:2014}. The long ago proposed unusual Rydberg system -- a charge He nanodroplet with an electron orbiting outside~\cite{Golov:1993} -- has finally been realized and characterized experimentally~\cite{LoginovPRL:2011,Lackner:2011}. 

The formation of dopant-He excited complexes has been probed for different species and using various photoionization techniques~\cite{Droppelmann:2004,Mudrich:2008,Giese:2012,Loginov:2014}. Although different formation mechanisms have been ascertained (direct association vs. tunneling), a detailed description is still evolving~\cite{Leino:2011,Vangerow:2014}. So is the dynamics of the collapse of a foam-like structure of magnesium atoms formed by aggregation inside He droplets when being irradiated by ultrashort light pulses~\cite{Goede:2013}. Femtosecond vibrational and electronic coherence spectroscopy has probed the decoherence and relaxation dynamics of molecules weakly bound to the He droplet surface~\cite{Mudrich:2008,Schlesinger:2010,Gruner:2011}. Using newly available ultrashort laser pulses, the exceptional behavior of doped He droplets when undergoing the transition into highly charged nanoplasmas have been explored theoretically and experimentally~\cite{KrishnanPRL:2011,Krishnan:2012}. 

Aside from doped He nanodroplets, new insight into the dynamics of pure He droplets subjected to electronic excitation has been gained owing to the advent of ultrafast sources of EUV radiation. A panoply of relaxation channels presents itself, including fluorescence emission~\cite{Haeften:2005}, the ejection of Rydberg atoms~\cite{Kornilov:2011} and zero-kinetic energy electrons~\cite{Peterka:2003}, intra- and inter-band relaxation~\cite{Buenermann:2012} and nanoplasma formation by collective autoionization~\cite{Ovcharenko:2014}. While most of these processes still lack a detailed theoretical description, with the current advances in computational tools and power accurate theoretical modeling of the dynamics of doped He nanodroplets including quantum effects is now within reach. 

Naturally, the most pertinent open question still pertains to the role of superfluidity and the quantum nature in various dynamical processes and observables such as spectral line shapes and shifts, the ejection dynamics of excited cations and neutrals in Rydberg states, the relaxation of rotational, vibrational and electronic excitations of dopants, and the dynamics of complex aggregation inside He droplets. The translational motion of atoms and molecules in nanometer-sized superfluid He droplets has recently been confirmed to proceed akin to moving macroscopic objects in bulk superfluid He, including the existence of a critical velocity for the undamped motion~\cite{Brauer:2013}. However, the fast damping of internal excitations of dopants by the He droplets, which is observed in the case of impulsive rotational~\cite{PentlehnerPRL:2013}, vibrational (see Fig.~\ref{fig:LiI}) and electronic excitation~\cite{Fechner:2012,Loginov:2012,Kautsch:2013} cannot definitely be correlated to the superfluid character of the He droplets so far. Likewise, further time-resolved experimental as well as theoretical studies are needed to elucidate the response of He droplets to the creation and excitation of ions in He droplets. Understanding on a microscopic level the peculiar properties of He droplets to act as an extremely efficient cooling agent on the one hand, and as a nearly nonperturbing frictionless superfluid on the other, remains an important challenge. In particular the potential of He droplets to serve as ``nano-cryo-reactors'' for studying the reaction dynamics of embedded species at ultralow temperatures needs to be further tested. Possibly, the emerging ultrafast multidimensional spectroscopic techniques~\cite{Jonas:2003} could be gainfully applied to attaining a more complete picture of the photodynamics of doped He nanodroplets. 

%The use of novel sources of light such as intense ultrashort laser pulses as well as EUV radiation emitted by high-harmonics generation and free-electron lasers opens up new opportunities, on the one hand, for studying fundamental properties of He droplets as a paradigm for simple atomic systems with bulk density at the nanoscale, on the other hand, to develop and refine new ionization based detection schemes for droplet-isolated species which allow to study fundamental properties and dynamics of larger molecular systems.

%The apparent fast migration of He$^*$ excitations to the surface of He droplets~\cite{Kornilov:2011,Buchta:2013} still lacks a conclusive picture. 
The vibronic dynamics of singly and multiply excited pure He droplets irradiated by EUV light~\cite{Kornilov:2011,Buchta:2013,Ovcharenko:2014} will be further studied using newly available radiation sources to probe the transition from atomic to condensed phase photodynamics of He. New theoretical approaches such as microscopic calculations of the electronic structure and dynamics of He clusters~\cite{Closser:2010,Closser:2014} may turn out to be instrumental for the interpretation of the experimental findings. Likewise, the so far unresolved ultrafast ignition dynamics of a nanoplasma inside He droplets, calls for further experiments exploiting few-cycle and attosecond laser pulses which are now spreading into many molecular and chemical physics laboratories.

The aggregation dynamics of dopant atoms in He droplets has recently been found to be driven by vortices -- quantized states of angular momentum in superfluid He droplets produced in the expansion of liquid He~\cite{Gomez:2012}. Vortices were traced by introducing Ag atoms, which clustered along the vortex lines, and appeared as track-shaped surface deposits in electron micrographs. Very recently, direct x-ray diffraction imaging using intense free-electron laser x-ray pulses has provided spectacular views of the structure of pure and doped droplets such as various non-spherical droplet shapes and regular patterns of vortices inside the droplets~\cite{Vilesov:2013}. Such direct imaging experiments, combined with femtosecond laser excitation and simultaneous electron and ion detection, will certainly provide complementary means to unravel the remaining riddles which He droplets still pose us.

\section{\label{sec:Acknowledgement}Acknowledgement}
We gratefully acknowledge fruitful discussions with T. Fennel, U. Saalmann, A. Heidenreich, M. Barranco, and with the members of the groups of O. Gessner and D. Neumark in Berkeley during the sabbatical of F. S.. Financial support by the Deutsche Forschungsgemeinschaft (DFG) is acknowledged.

%\bibliography{HeDropletPI_Bib}% Produces the bibliography via BibTeX.

\newpage

\newpage

\newpage

\newpage

\newpage

\newpage

\newpage

\newpage

\newpage

\newpage

\newpage

\newpage

\end{document}